\documentclass[reprint,amsmath,amssymb,aps,eqsecnum,showpacs,prx]{revtex4-1}

\usepackage{graphicx}
\usepackage{bm}
\usepackage{epsfig}
\usepackage{amssymb}
\usepackage{amsfonts}
\usepackage{braket}
\usepackage{color}
\usepackage{epstopdf}
\usepackage{hyperref}
\usepackage{float}
\usepackage{bibentry}
\usepackage[caption=false]{subfig}

\newcommand{\ba}{\begin{eqnarray}}
\newcommand{\ea}{\end{eqnarray}}
\newcommand{\bd}{\begin{displaymath}}

\newcommand{\nn}{\nonumber \\}

\graphicspath{{figures/}}

\begin{document}
\title{Classification of trivial spin-1 tensor network states on a square lattice}
\author{Hyunyong Lee}
\author{Jung Hoon Han}
\affiliation{Department of Physics, Sungkyunkwan University, Suwon 16419, Korea}
\date{\today}

\begin{abstract}
Classification of possible quantum spin liquid (QSL) states of interacting spin-1/2's in two dimensions has been a fascinating topic of condensed matter for decades, resulting in enormous progress in our understanding of low-dimensional quantum matter. By contrast, relatively little work exists on the identification, let alone classification, of QSL phases for spin-1 systems in dimensions higher than one. Employing the powerful ideas of tensor network theory and its classification, we develop general methods for writing QSL wave functions of spin-1 respecting all the lattice symmetries, spin rotation, and time reversal with trivial gauge structure on the square lattice. We find $2^5$ distinct classes characterized by five binary quantum numbers. Several explicit constructions of such wave functions are given for bond dimensions $D$ ranging from two to four, along with thorough numerical analyses to identify their physical characters. Both gapless and gapped states are found. The topological entanglement entropy of the gapped states are close to zero, indicative of topologically trivial states. In $D=4$, several different tensors can be linearly combined to produce a family of states within the same symmetry class. A rich ``phase diagram" can be worked out among the phases of these tensors, as well as the phase transitions among them. Among the states we identified in this putative phase diagram is the plaquette-ordered phase, gapped resonating valence bond phase, and a critical phase. A continuous transition separates the plaquette-ordered phase from the resonating valence bond phase. 
\end{abstract}

\maketitle

\section{Introduction}

In laymen's terms, quantum spin liquid (QSL) refers to the ground state of some low-dimensional spin Hamiltonian that lacks long-range magnetic order. Most model Hamiltonians in one dimension (1D) would have QSL ground state according to this simple criterion, by virtue of the Mermin-Wagner theorem.  Although instances of their existence are still rare in higher dimensions, the variety of possible QSL phases in theory has grown enormously since its inception in the context of frustrated quantum magnets\,\cite{balents10}. 

A remarkable observation due to Haldane in the early 80s\,\cite{haldane83} is that a sharp distinction can be drawn among the 1D QSLs depending on whether a mass gap exists in the excitation spectrum. The distinction is in turn dictated by the size of the constituent spin being integer or half-integer, and the impact it has on the corresponding global Berry phase behavior. In this regard one may say the earliest effort at classification of spin liquid states is rooted in field-theoretical approach, with emphasis on the sensitivity of the Berry phase on the spin size. The condition imposed by the spin size is by no means definitive, as one can easily write down both gapless and gapped QSL models using spin-1/2 objects. A well-known example along this line is the $J_1 -J_2$ spin-1/2 chain in which even a tiny bond modulation produces a phase transition of the gapless phase into the gapped one\,\cite{kohmoto92, hida92}. Also for spin-1 chains where a gap was predicted, a whole myriad of different phases, some gapped and some not, can be found in an interesting bi-linear bi-quadratic extension of the Heisenberg spin-1 Hamiltonian\,\cite{lauchli06}. A modern effort at classifying phases of 1D QSL is based on the idea of ``protection by symmetries", as summarized in the work of Chen et al.\,\cite{chen11} and references therein. 

Haldane extended his Berry phase analysis to spin models in two dimensions\,\cite{haldane88}, where it was pointed out that certain Skyrmion proliferation processes are allowed or forbidden depending on the spin size $S$. On a square lattice, for instance,  one encounters constructive interference of instanton processes favoring the Skyrmion proliferation, leading to the fully gapped ground state for $S=2$\,\cite{haldane88,aklt88}. For lesser spins $S<2$, we know that $S=1/2$ Heisenberg antiferromagnet on a square lattice has a long-range ordered antiferromagnetic ground state, which becomes disordered by the addition of sufficiently strong diagonal exchange interaction\,\cite{jiang12}. Much like the construction in one dimension, the character of the ground state depends on the details of the Hamiltonian as well as the size of the spin itself. Even more than in one dimension, complete and reliable classification of QSL in two dimensions requires schemes going well beyond the Berry phase picture. 

More recently, an ambitious new paradigm to classify phases in low-dimensional quantum magnets and construct their explicit wave functions has been launched under the theme of tensor network (TN) approach. In deep contrast to earlier approaches to QSL based on slave-particle theories and Gutzwiller-projected wave function studies\,\cite{wen04book}, the TN-based ideas set out by the assumption that the relevant low-energy wave functions can be constructed as the tensor network form, with little attention devoted (at least for now) to the identification of the Hamiltonian for which such tensor would be the ground state.  

Literature on the tensor network theory of the QSL has grown vastly over the past decade, and we refer the readers to Ref.\,\onlinecite{shenghan15A} for an exhaustive list of relevant papers. An important line of efforts has been aimed at clarifying the role of symmetries in constructing tensor wave functions. Loosely speaking, symmetry properties one wishes to impose on the many-body wave function are implemented explicitly at the level of the local tensor\,\cite{pollmann10,vidal11,vidal12,shenghan15A,shenghan15B}, thus ensuring that the global many-body state obtained from contraction of local tensors obey the required symmetries. 

For one dimension, where the tensor network state goes by another name MPS (matrix product states), symmetry constraints on the local MPS matrix has been understood by several authors\,\cite{pollmann10,chen11}. An analogous effort in two dimensions is much more challenging, both due to bigger symmetries of the crystal structure and the greater size of the local tensor with three (honeycomb), four (kagome, square), or six (triangular) bond indices instead of two (one dimension), for each physical spin index in a local tensor. An effort at symmetry classification of the tensor has been paralleling the development of the tensor network theory itself. A most recent, ambitious take on the TN symmetry classification is the work of Jiang and Ran (JR)\,\cite{shenghan15A}. 

The JR approach is, in essence, a successful implementation of the earlier projective symmetry group (PSG) idea\,\cite{wen04book}, re-engineered to treat the symmetry properties of the local tensor in a TN wave function in a projective manner. With these ideas, JR were able to predict a certain number of distinct classes of TN wave functions for spin-1/2 QSL with $\mathbb{Z}_2$ symmetry on a kagome lattice\,\cite{shenghan15A}. Following their initiative, Ref.\,\onlinecite{shenghan15B} constructed a symmetric TN wave function for interacting spin-1/2's on a honeycomb lattice which is devoid of topological order. 

As in the PSG formulation, the concept of invariant gauge group or IGG for short, plays a central role in the tensor network theory, in both classifying quantum phases and identifying the nature of excitations for each quantum ground state. It is expected, although not rigorously proven, that tensor networks with a trivial IGG will be devoid of topological order and do not allow fractionalized excitations. In this paper we are mostly devoted to tensor network states with trivial IGG, for physical spin-1. At first sight it appears that TN states with trivial IGG are in line with the recent notion of symmetry-protected trivial (SPT) phases of quantum matter\,\cite{senthil15}, which is also devoid of topological order by construction. A key difference, however, is that SPT states possess protected gapless states at the boundary, while no such guarantee exists for TN states with trivial IGG. It is not even required that TN states consistent with a certain set of symmetry requirements should possess an energy gap. As we will show by explicit examples in this paper, the TN states can as easily be gapless and critical, as it can be gapped. The tensor network theory is a flexible way to construct a variety of QSL states, constrained only by the symmetry properties we impose. In this sense, the approach initiated in Refs.\,\onlinecite{shenghan15A,shenghan15B} and adopted here is similar to the notion of ``fragile Mott insulator", which is also a symmetry-based classification scheme of Mott insulators lacking in topological order\,\cite{yao10}. 


Compared to the spin-1/2 model on various two-dimensional lattices where vast literature exists, including the two references mentioned above\,\cite{shenghan15A,shenghan15B}, relatively little attention has been given to understanding possible phases and their dynamics for spin-1 models\,\cite{jiang09,wei14,wang15,gong15}. As an exception we note the DMRG work of Jiang et al. which studied the phase diagram of the spin-1 $J_1 -J_2$ Hamiltonian on a square lattice\,\cite{jiang09}. Indications of the spin liquid phase for $J_2/J_1 \approx 0.5$, sandwiched between $(\pi,\pi)$- and $(\pi,0)$-ordered antiferrmagnetic ground states, were given. 
Using the tensor network idea, Li et al. constructed the so-called resonating AKLT loop (RAL) state on the square lattice sitting at the critical phase\,\cite{wei14}, while a gapped, featureless paramagnetic $S=1$ tensor wave function was constructed in Ref. \onlinecite{zaletel16}. Both constructions were devoid of any topological order\,\cite{wei14,zaletel16}. The possibility of deconfined quantum critical phase transition in the spin-1 model on the honeycomb lattice was raised in Ref. \onlinecite{tao12}.

Despite these instances of exemplary constructions, a formal and thorough classification of possible TN states for $S=1$ is still lacking in any lattice geometry in two dimensions. In the TN language the constructed wave function in Ref.\,\onlinecite{zaletel16} has the bond dimension $D=2$ and the one in Ref.\,\onlinecite{wei14}, $D=3$. At the moment we do not have any idea whether these constructions represent a generic spin-1 phase on the square lattice that respect all the symmetries of the lattice and two internal ones, namely time reversal and spin rotation, or random examples out of a vast pool of $S=1$ QSL states. To address this question successfully, it is imperative to carry out a methodical classification of possible symmetric TN wave functions following the scheme put forward by JR\,\cite{shenghan15A}. The outcome of our analysis are organized according to the bond dimension of the virtual space ranging from $D=2$ to $D=4$. For each bond dimension $D$ we carry out a thorough symmetry analysis of possible tensor forms. As a result of our investigation we arrive at the one proposed in Ref.\,\onlinecite{zaletel16} as the sole tensor compatible with all symmetry requirements for $D=2$. Our construction of the $D=3$ tensor is new, and consists of two tensors with mutually different internal quantum numbers. Examination of their physical properties through correlation function analyses reveal that one is a fully gapped state, and the other represents a critical state. A much richer picture emerges at $D=4$. There are three distinct tensors within a given symmetry class, which can be combined arbitrarily to produce a family of tensors still within the same symmetry class. We can then map out a ``phase diagram" within this family of tensor states, find several distinct phases, and study phase transitions among them. 

Throughout this paper we are restricted to tensors with a trivial IGG whose notion in the context of tensor network is explained in Refs. \onlinecite{swingle10,shenghan15A}. Classification of symmetric tensors with nontrivial IGG's, for instance IGG=$\mathbb{Z}_2$, requires a completely independent analysis which we defer to a different publication. Due to the restriction to trivial IGG the tensor states we construct are expected to belong to the ``trivial" states devoid of topological order. Nevertheless it can happen, as with examples found in Refs. \onlinecite{shenghan15A,shenghan15B}, that the state that passed the filtering of symmetry and IGG analyses possess a new, ``emergent symmetry"  with non-trivial IGG. When that happens we will write out the nature of the emergent IGG and its origin. Section\,\ref{sec:class} is devoted to the formal theory of classification of symmetric tensors for spin-1 on a square lattice. It is followed in Sec.\,\ref{sec:d23} by explicit constructions of symmetry-compatible tensors for bond dimensions $D=2,3$ and analyses of their correlation functions. Since $D=4$ symmetric tensors are much richer and an interesting ``phase diagram" can be constructed in the space of tensor wave functions, we devote Sec.\,\ref{sec:d4} to both the derivation of the $D=4$ tensors, their correlation functions, and the construction of the phase diagram. In the last section we try to give some perspective on the significance of our analyses and summarize. Throughout the paper we use the terms PEPS (projected entangled pair states) and TN (tensor network) interchangeably. 

\section{Classification of spin-1 symmetric PEPS on square lattice with trivial IGG}
\label{sec:class}

\begin{figure}\includegraphics[width=0.5\textwidth]{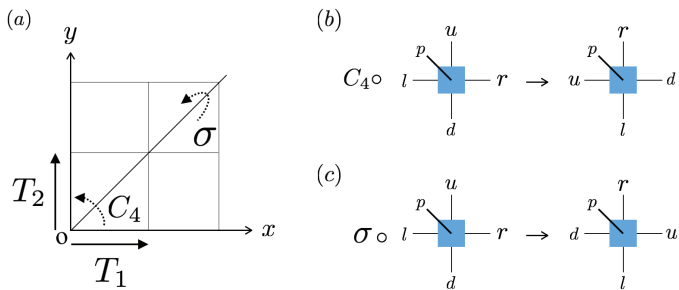}
	\caption{(a) Schematic figure of lattice symmetry operations on a square lattice. $T_{1,2}$: translation along the $x,y$-direction, $C_4$: $\pi/2$ spatial rotation about the origin, $\sigma$: reflection about the $y=x$ axis. (b) Site tensor transformation under the $C_4$ operation. (c) Site tensor under the $\sigma$ operation. Virtual legs are labeled by $l,r,u,d$, and the physical leg by $p$.}
	\label{fig:schematic}
\end{figure}

We give a brief account in this section of the general classification theory for spin-1 tensor network wave functions on a square lattice. Explicit constructions of TN wave functions are delegated to the next two sections, where their physical properties will be delineated by extensive numerical analyses.  Much of the philosophy behind the classification ideas being followed in this paper has been laid out before\,\cite{shenghan15A,shenghan15B}. Most of the details of the classification work specific to spin-1 can be found in the Appendix\,\ref{appendix:classification}. Note that a formal and rigorous classification of spin-1 PEPS wave functions has been lacking in the literature. All previous classifications of PEPS were in regard to physical spin-1/2's\,\cite{wen02,wang06,ran11,yang15,shenghan15A,shenghan15B}.

Throughout the paper, {\it symmetric} PEPS refers to those TN wave functions consistent with all of the imposed symmetries. In our case they are two internal symmetries (spin rotation and time reversal) and all point-group symmetries of the square lattice. As is usual in the PEPS construction, we introduce two different tensors, site and bond tensors denoted $\mathbf{T}$ and $\mathbf{B}$ respectively. Symmetry requirements impose the following algebraic conditions on them\,\cite{swingle10, shenghan15A}

\begin{align}
	\mathbf{T} &= \Theta_R W_R R \circ \mathbf{T},\nn
	\mathbf{B} &= W_R R \circ \mathbf{B} .
	\label{eq:sym_peps}
\end{align}
$W_R$ and $\Theta_R$ respectively indicate the gauge transformation matrix and the U(1) phase factor associated with each symmetry operation $R$\,\cite{shenghan15A,shenghan15B}. $R\circ\mathbf{T}$ and $R\circ \mathbf{B}$ symbolically express the symmetry operations on the tensors. 

To orient readers unfamiliar with the tensor network terminology, one can view the two-index bond tensor $\mathbf{B}$ as a $D\times D$ matrix residing at the link between adjacent sites of the lattice. The site tensor $\mathbf{T}$ has one physical index, of dimension 3 (since we are considering spin-1), and four ``virtual indices" each with dimension $D$. Each virtual index of the site tensor is to be contracted with one ``leg" of the bond tensor. The size of the virtual Hilbert space $D$ is called the bond dimension. 

On a square lattice, spatial symmetry is fully defined by the four generators $\{T_1,\,T_2, C_4, \sigma \} $ corresponding to translations along $x$- and $y$-directions ($T_1 , T_2$), $\pi/2$-rotation around the origin ($C_4$), and reflection about the $y=x$ axis ($\sigma$). Figure\,\ref{fig:schematic} schematically depicts the action of each generator on a square lattice and the site tensor. These four generators completely define the space group of the square lattice through the following commutative relations

\begin{align}
	T_2^{-1} T_1^{-1} T_2 T_1 &= e,\nn
	C_4^{-1} T_1 C_4 T_2 &= e,\nn
	C_4^{-1} T_2 C_4 T_1^{-1} &= e,\nn
	(C_4)^4 &= e,\nn
	\sigma^{-1} T_2^{-1} \sigma T_1 &= e,\nn
	\sigma^{-1} T_1^{-1} \sigma T_2 &= e,\nn
	(\sigma)^2 &=e,\nn
	\sigma^{-1} C_4 \sigma C_4 &= e.
	\label{eq:space_group}
\end{align}
Since the time reversal\,($\mathcal{T}$) and SU(2) spin rotation\,($U_{\theta\vec{n}}$) operations commute with each other as well as with all space symmetry operations, the following relations must hold as well: 

\begin{align}
	g^{-1} \mathcal{T}^{-1} g \mathcal{T} &= e,\,\,\forall g=T_1,T_2,C_4,\sigma,\nn
	g^{-1} U_{\theta\vec{n}}^{-1} g U_{\theta\vec{n}} &= e,\,\,\forall g=T_1,T_2,C_4,\sigma,\mathcal{T}.
	\label{eq:time_spin_rotation}
\end{align}

Readers may wonder why we are imposing SU(2) symmetry on spins when its size $S=1$ seems to call for SO(3) symmetry instead. Although the physical spin of interest in this paper is spin-1, the internal spin rotation operation acts on the virtual spin degrees of freedom as well as on physical spins, and we are allowing the possibility of half-integer spins for the virtual spins. Hence it is appropriate to speak of the SU(2), rather than SO(3), spin rotation symmetry. We also consider the following two relations

\begin{align}
	\mathcal{T}^2 = e,\,\,\,\,\,\, U_{\theta=2\pi} = e,
	\label{eq:time2_spin_rotation}
\end{align}
to hold when acting on the TN wave functions.

The commutative relations (\ref{eq:space_group}) and (\ref{eq:time_spin_rotation}) provide algebraic equations for $W_R$ and $\Theta_R$, where $R$ ranges over both spatial and internal symmetry operations $R\in \{T_1,T_2,C_4,\sigma,\mathcal{T},U_{\theta\vec{n}} \}$\,\cite{shenghan15A}. Solving such equations provides the following ``solutions" for $W_R$'s and $\Theta_R$'s: 

\begin{align}
	& W_{T_{1,2}}(x,y,i) = \mathbb{I}_D ,\;\;\; \Theta_{T_{1,2}}(x,y) = 1,\nn
	& W_{C_4}(x,y,i) = \mathbb{I}_D ,\;\;\; \Theta_{C_4}(x,y) = \theta_{C_4},\nn
	& W_{\sigma}(x,y,i) = \bigoplus_{i=1}^M \left( \widetilde{w}_{\sigma}^i \otimes \mathbb{I}_{2S_i+1} \right),\;\;\; \Theta_{\sigma}(x,y) = \theta_{\sigma},\nn
	& W_{\mathcal{T}}(x,y,i) = \bigoplus_{i=1}^M \left( \widetilde{w}_{\mathcal{T}}^i \otimes e^{i\pi S_i^y } \right),\;\;\; \Theta_{\mathcal{T}}(x,y) = 1,\nn
	& W_{\theta}(x,y,i) = \bigoplus_{i=1}^M \left( \mathbb{I}_{d_i} \otimes e^{i\theta \vec{n} \cdot \vec{S}_i} \right),\;\;\; \Theta_{\theta}(x,y) = 1 .
	\label{eq:sol_classification}
\end{align}
Subscripts under $\mathbb{I}$'s give the dimension of each identity matrix. 

The Hilbert space of each virtual leg is spanned by $M$ different species of spins, i.e. $\mathbb{V}_{\vec{S}_i}$ with $i$ ranging from 1 to $M$ and $\vec{S}_i$ is a virtual spin of size $S_i$. For each spin species $\vec{ S}_i$, we further introduce the ``flavor degeneracy", by which one means that there are $d_i$ identical copies of the same spin $\vec{ S}_i$. The flavor Hilbert space for spin $\vec{ S}_i$ is denoted $\mathbb{D}_i$, and the entire Hilbert space for the virtual leg becomes
\begin{align}
	\mathbb{V}_{\rm virtual} = \bigoplus_{i=1}^M (\mathbb{D}_i \otimes \mathbb{V}_{\vec{S}_i}).
	\nonumber
\end{align}
The bond dimension is the sum, 

\ba D=\sum_{i=1}^M (2S_i +1) \cdot d_i. \nonumber \ea
The flavor degeneracy $d_i$ is introduced for completeness, but usually one can realize symmetric tensors within $d_i = 1$ subspace having no flavor degeneracy. For instance one has the freedom to realize $D=4$ bond dimension with $S=1/2$ and $d=2$, or with a single $S=3/2$ and $d=1$. The choice is one of convenience, not of principle. 

Each $W_R$ is defined on one of the four bonds $i \in \{l,r,u,d\}$ emanating from a given site $(x,y)$: $W_R (x,y,i)$. Each $\Theta_R$ is defined on a given site: $\Theta_R (x,y)$.  
Binary integer values $\pm 1$ are assigned for each of the two ``quantum numbers" $\theta_{C_4}$ and $\theta_\sigma$ introduced in Eq. (\ref{eq:sol_classification}). The two matrices $\widetilde{w}_{\mathcal{T}}^i$ and $\widetilde{w}_{\sigma}^i$ acting on the flavor space are both $d_i$-dimensional matrices satisfying 

\ba 
\widetilde{w}_{\mathcal{T}}^i (\widetilde{w}_{\mathcal{T}}^i)^* & =&  \chi_{\mathcal{T}} \cdot \chi_{\theta=2\pi} \mathbb{I}_{d_i}, \nn (\widetilde{w}_{\sigma}^i)^2 & =&  \mathbb{I}_{d_i} ,\nn 
W_{\sigma}^{-1} (W_{\mathcal{T}}^{-1})^* W_{\sigma}^* W_{\mathcal{T}}^* &=& \chi_{\sigma \mathcal{T}}\mathbb{I}_{d_i} .
\label{eq:constraints}
\ea
For the simplest case with $d_i = 1$ these matrices reduce to numbers, e.g. $|\widetilde{w}_{\mathcal{T}}^i |^2 =  \chi_{\mathcal{T}} \cdot \chi_{\theta=2\pi}  = 1$, and $(\widetilde{w}_{\sigma}^i)^2 = 1$. Origin of these relations can be looked up in the Appendix, but its knowledge is not essential for comprehending the discussions in the main body of the paper. There are three additional binary quantum numbers introduced by (\ref{eq:constraints}), $\chi_{\mathcal{T}}, \chi_{\theta=2\pi}, \chi_{\sigma\mathcal{T}}$. 

There can be at most $2^5$ different {\it classes} of wave functions distinguished by five sets of binary quantum number $\{ \theta_{C_4}, \theta_{\sigma}, \chi_{\mathcal{T}}, \chi_{\sigma\mathcal{T}},\chi_{\theta=2\pi} \}$ in our classification framework. Among these numbers, $\chi_{\theta=2\pi}$ deserves a special mention as it plays an important role in constraining the virtual spin sizes. The second relation in Eq.\,\eqref{eq:time2_spin_rotation} implies $W_{\theta=2\pi} = \chi_{\theta=2\pi}\mathbb{I}_D$  (see Appendix for proof of this statement). From the representation of $W_{\theta}$ in Eq.\,\eqref{eq:sol_classification} one can read off that all virtual spins $\vec{S}_i$ must transform identically, either as half-odd integers with $\chi_{\theta=2\pi} = -1$ or as integers with $\chi_{\theta=2\pi} = +1$. As an immediate and powerful consequence, it is impossible to use ``mixed" representations with both half-integer and integer virtual spins. The RAL construction adopted in Ref.\,\cite{wei14} employed a mixed spin-0 and spin-1/2 virtual spin representation and lies outside of our classification framework. 

\section{spin-1 PEPS with bond dimensions $D=2,3$ }
\label{sec:d23}

In this section we provide explicit constructions of symmetic TN wave functions for the bond dimensions $D=2$ and $D=3$. The $D=2$ construction is only possible with $M=1$, $d=1$ (trivial flavor space) and $S=1/2$ (virtual spin-1/2). There is only one $D=2$ construction consistent with all the symmetry requirements, and this turns out to be equivalent to the ansatz proposed in Ref.\,\cite{zaletel16}. We give a brief account of the $D=2$ construction.

Firstly, the bond tensor should fuse two virtual spin-1/2's into a spin singlet, 
\begin{align}
	& [\mathbf{B}]_{lr} = C_{\frac{1}{2},m_{l};\frac{1}{2},m_{r}}^{0,0} , \nn
	& [\mathbf{B}]_{ud} = C_{\frac{1}{2},m_{u};\frac{1}{2},m_{d}}^{0,0}. 
\end{align}
We introduced Clebsch-Gordon (CG) coefficients $C_{S_1,m_1;S_2,m_2}^{S_3;m_3}$ for fusing two spins\,($S_1$, $S_2$) into a resulting spin\, $S_3$, and $m_i$ denotes $S^z$ quantum numbers satisfying $m_3 = m_1 + m_2$. Site tensor also must be a spin singlet, and throughout this article, we use the following {\it fusion tree}\,\cite{vidal12} to obtain the spin singlet site tensor: (1) fuse two spins living on the left and right (or up and down) legs into an intermediate spin, (2) fuse two intermediate spins into the virtual spin-1, (3) fuse the resulting virtual spin-1 and the physical spin-1 into a spin singlet. Particularly, with $D=2$, the Hilbert space of the site tensor can be decomposed as

\begin{align}
	\mathbb{V}_{\mathbf{T}_2}^{S=0} & \cong  \left(\mathbb{V}_{\frac{1}{2}\frac{1}{2}}^{S=1} \otimes \mathbb{V}_{\frac{1}{2}\frac{1}{2}}^{S=1} \otimes \mathbb{V}_{11}^{S=1} \otimes \mathbb{V}_{11}^{S=0}	\right)\nn
	& \oplus \left(\mathbb{V}_{\frac{1}{2}\frac{1}{2}}^{S=1} \otimes \mathbb{V}_{\frac{1}{2}\frac{1}{2}}^{S=0} \otimes \mathbb{V}_{10}^{S=1} \otimes \mathbb{V}_{11}^{S=0} \right)\nn
	& \oplus \left(\mathbb{V}_{\frac{1}{2}\frac{1}{2}}^{S=0} \otimes \mathbb{V}_{\frac{1}{2}\frac{1}{2}}^{S=1} \otimes \mathbb{V}_{01}^{S=1} \otimes \mathbb{V}_{11}^{S=0} \right)\nn
	& \cong \mathbb{V}_{\mathbf{T}_2}^{(11)} \oplus  \mathbb{V}_{\mathbf{T}_2}^{(10)} \oplus  \mathbb{V}_{\mathbf{T}_2}^{(01)}, \label{eq:fusion-D-2}
\end{align} 
representing three distinct fusions processes giving rise to the total singlet site tensor. We defined $\mathbb{V}_{\mathbf{T}_D}^{S=0}$ as the Hilbert space of the spin singlet site tensor $\mathbf{T}_D$ with the bond dimension $D$, $\mathbb{V}_{S_1 S_2}^{S_3}$ as the fusion space fusing spins $S_1$ and $S_2$ into $S_3$\,\cite{vidal12,shenghan15A}. The fusion products in each line of the above equation must be read from left to right. The three different ``trees" labeled as 
$\mathbb{V}_{\mathbf{T}_2}^{(11)}, \mathbb{V}_{\mathbf{T}_2}^{(10)}, \mathbb{V}_{\mathbf{T}_2}^{(01)}$ are distinguished by the values of intermediate virtual spins. For instance $\mathbb{V}_{\mathbf{T}_2}^{(11)}$ means that combining the left and right virtual spin-1/2's give the intermediate spin-1, and the top and bottom virtual spins also giving the intermediate spin-1. 
Site tensors incorporating the above fusion rules can be constructed using the CG coefficients.  For instance, 

\begin{align}
	& [\mathbf{T}^{(10)}_2]_{lrud}^p = C_{\frac{1}{2},m_l;\frac{1}{2},m_r}^{1,n_1} C_{\frac{1}{2},m_u;\frac{1}{2},m_d}^{0,0} C_{1,n_1;0,0}^{1,n_2}   C_{1,m_p;1,n_2}^{0,0} \nn
	& [\mathbf{T}^{(01)}_2]_{lrud}^p = C_{\frac{1}{2},m_l;\frac{1}{2},m_r}^{0,0} C_{\frac{1}{2},m_u;\frac{1}{2},m_d}^{1,n_1} C_{0,0;1,n_1}^{1,n_2}   C_{1,m_p;1,n_2}^{0,0}.
\end{align}

Since the two fusion processes $\mathbb{V}_{\mathbf{T}_2}^{(10)}$ and $\mathbb{V}_{\mathbf{T}_2}^{(01)}$ involve intermediate spins which differ between left-right and up-down addition of spins, one might suspect the $C_4$ symmetry is broken. Indeed this is the case for the ``bare" tensor, and one must apply symmetrization process to restore the $C_4$ symmetry:

\begin{align}
	\mathbf{T'}^{(10)}_2 =  \sum_{n=0}^3 (\theta_{C_4}^{-1} C_4)^n \circ \mathbf{T}^{(10)}_2,\nn
	\mathbf{T'}^{(01)}_2 =  \sum_{n=0}^3 (\theta_{C_4}^{-1} C_4)^n \circ \mathbf{T}^{(01)}_2.\nonumber
\end{align}
It turns out, however, that both tensors $\mathbf{T'}^{(10)}_2, \mathbf{T'}^{(01)}_2$ become zero after the symmetrization. The only possible symmetric site tensor comes from the fusion tree $\mathbb{V}_{\mathbf{T}_2}^{(11)}$, with the explicit CG coefficients given by 

\begin{align}
	& [\mathbf{T}^{(11)}_2]_{lrud}^p = C_{\frac{1}{2},m_l;\frac{1}{2},m_r}^{1,n_1} C_{\frac{1}{2},m_u;\frac{1}{2},m_d}^{1,n_2} C_{1,n_1;1,n_2}^{1,n_3}   C_{1,m_p;1,n_3}^{0,0} \nn
	&= (-1)^{2-m_p} \delta_{m_p,-n_3} C_{\frac{1}{2},m_l;\frac{1}{2},m_r}^{1,n_1} C_{\frac{1}{2},m_u;\frac{1}{2},m_d}^{1,n_2} C_{1,n_1;1,n_2}^{1,n_3}   \nn
	&= (-1)^{m_p} C_{\frac{1}{2},m_l;\frac{1}{2},m_r}^{1,n_1} C_{\frac{1}{2},m_u;\frac{1}{2},m_d}^{1,n_2} C_{1,n_1;1,n_2}^{1,m_p} .
	\label{eq:d2_sym_tensor}
\end{align}
Reading the product of CG coefficient from left to right, one can see how the fusion rules of Eq. (\ref{eq:fusion-D-2}) are being realized.  All indices except $m_p$ (physical index) and $m_l, m_r, m_u, m_d$ (virtual indices) in Eq. (\ref{eq:d2_sym_tensor}) are being summed over. The physical index $p=1,2,3$ refers to the three $S^z$ basis states $|+\!1\rangle,\,|0\rangle,\,|-\!1\rangle$ with $m_1 = 1, m_2 = 0$ and  $m_3 =-1$. Redefining the physical spin basis, i.e. $|+\!1\rangle\!\rightarrow\! -|+\!1\rangle$ and $|-\!1\rangle \!\rightarrow\! -|-\!1\rangle$, the site tensor can be recast as

\begin{align}
	& [\mathbf{T}^{'(11)}_2]_{lrud}^p = C_{1,n_1;1,n_2}^{1,m_p} C_{\frac{1}{2},m_u;\frac{1}{2},m_d}^{1,n_2} C_{\frac{1}{2},m_l;\frac{1}{2},m_r}^{1,n_1}, \nonumber
\end{align}
in complete agreement with the tensor ansatz suggested in Ref.\,\cite{zaletel16}. It can be easily shown that only a single class characterized by 

\ba (\theta_{C_4},\theta_{\sigma}, \chi_{\mathcal{T}}, \chi_{\sigma\mathcal{T}}, \chi_{\theta=2\pi})=(-1,-1,+1,+1,-1) \nonumber \ea
is realizable with $D=2$, and all other ways to fuse a single spin-1 and 4 spin-1/2's to have the spin singlet are forbidden by lattice symmetries. Therefore, $\mathbf{T}_2^{(11)}$\,[Eq.\,\eqref{eq:d2_sym_tensor}] is the only tensor respecting all lattice symmetries, time-reversal and SU(2) spin rotation symmetries leading to the featureless quantum paramagnet\,\cite{zaletel16} at $D=2$.   

In the end, our procedures end up with a re-derivation of the known construction for $D=2$. The general strategy outlined here, however, will repeatedly appear in all subsequent constructions of symmetric tensors in higher bond dimensions and is worth carefully laying out. The quantum number aspect of the constructed tensor was not discussed before. Indeed, one of the main advantages of our elaborate classification scheme is in its power to assign quantum numbers $\theta_{C_4}$ through $\chi_{\theta=2\pi}$, as soon as the tensor construction is completed. Such quantum numbers do not play much role when there is only one possible tensor in a given bond dimension, such as the case with $D=2$. Shortly we will find that this is no longer the case for $D\ge 3$. There are two different symmetric site tensors one can obtain at $D=3$ from engineering of the CG coefficients. With the aid of quantum number characterization one can show that they have different sets of quantum numbers. Furthermore, their physical characters are completely different.

\begin{figure}\includegraphics[width=0.5\textwidth]{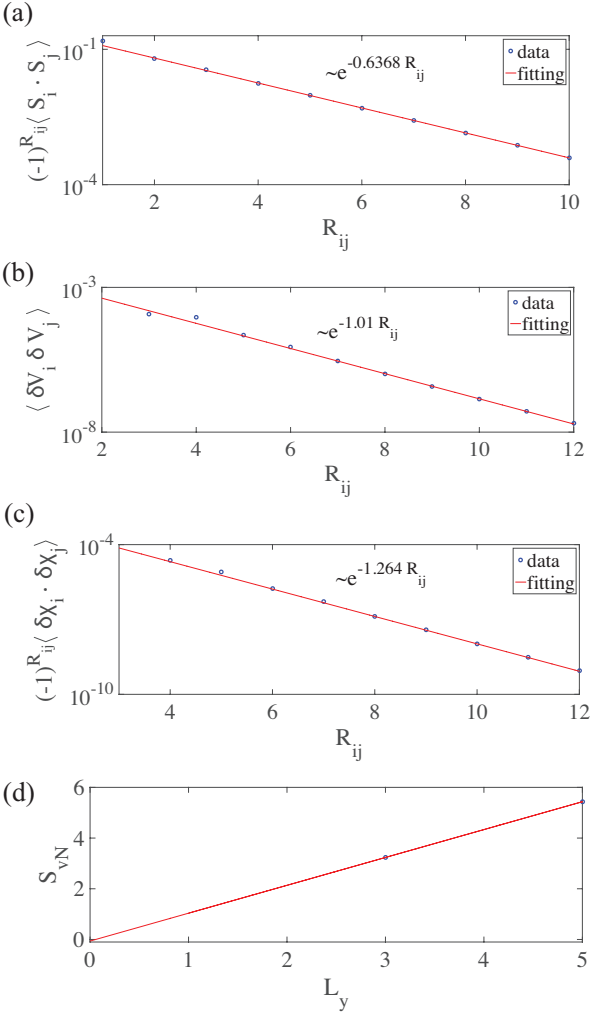}
	\caption{ (a) Spin, (b) dimer, (c) vector chirality correlators and (d) entanglement entropy\,(EE) for the PEPS made of $\mathbf{T}_3^{(11)}$. All correlations are measured with the finite-size PEPS of dimension $L_x\times L_y$. Measurement distance $R_{ij}$ is chosen as $L_x=L_y=L=3R_{ij}$ for successively increasing linear size $L$. Numerical data\,(blue circles) are fitted very well to exponential functions\,(red solid line). Best-fit exponential functions are given for each figure. EE is measured on a cylinder geometry\,(periodic boundary condition along $y$-direction) reaching the limit of $L_x\rightarrow \infty$, and the topological entanglement entropy is extracted from the fitting, $S_{vN}(L_y=0)=0.06$.}
	\label{fig:correlator_int_spin1}
\end{figure}
\begin{figure}\includegraphics[width=0.5\textwidth]{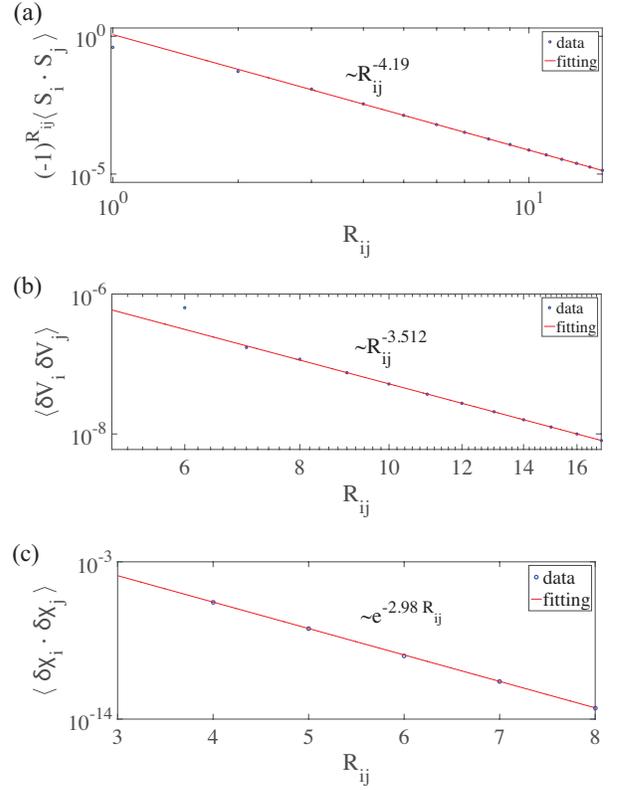}
	\caption{(a) Spin, (b) dimer, and (c) vector chirality correlators for PEPS made of $\mathbf{T}_3^{(22)}$. The same system size is employed as in Fig.\,\ref{fig:correlator_int_spin1}. Numerical results\,(blue circle) are fitted to algebraic (spin, dimer) or exponential (spin chirality) functions\,(red solid line). Best-fit power-law or exponential functions are shown for each figure.}
	\label{fig:correlator_int_spin2}
\end{figure}

Let us move to the more challenging case $D=3$. As explained in the previous section, we are not allowed to mix integer and half-integer spins in the virtual spin space. Thus, constructing a spin singlet site tensor with bond dimension $D=3$ is only possible if we assign a spin-1 Hilbert space for the virtual legs ($\chi_{\theta=2\pi}=+1$). It means that the degeneracy flavor space can only represent a Kramers singlet\,($\chi_{\mathcal{T}}=+1$) with dimension $d=1$. It also follows, from Eq. (\ref{eq:constraints}), that $\widetilde{w}_{\sigma} =\widetilde{w}_{\mathcal{T}} = 1$, as well as $\chi_{\sigma\mathcal{T}}=+1$. In this sense, we guess that the RAL state suggested in Ref.\,\cite{wei14}, where spin-0 and spin-1/2 are accommodated on virtual leg\,($D=3$), may have an emergent $Z_2$ or U(1) IGG since only the non-trivial IGG allows us to have mixture of integer and half-integer spins on virtual legs according to our classification. 

The remaining two quantum numbers $\theta_{C_4}$ and $\theta_{\sigma}$ and not yet constrained. Similar to $D=2$, the bond tensor is constructed by fusing two virtual spin-1's into a singlet, 

\begin{align}
	& [\mathbf{B}]_{lr} = C_{1,m_{l}; 1,m_{r}}^{0,0}, \nn & [\mathbf{B}]_{ud} = C_{1,m_{u}; 1,m_{d}}^{0,0}.
\end{align}
As regards the site tensor for $D=3$, we can consider the following two fusion rules of physical and virtual spins

\begin{align}
 	\mathbb{V}_{\mathbf{T}_3}^{(11)} &\cong \mathbb{V}_{11}^{S=1} \otimes \mathbb{V}_{11}^{S=1} \otimes \mathbb{V}_{11}^{S=1} \otimes \mathbb{V}_{11}^{S=0},\nn
 	 \mathbb{V}_{\mathbf{T}_3}^{(22)} &\cong \mathbb{V}_{11}^{S=2} \otimes \mathbb{V}_{11}^{S=2} \otimes \mathbb{V}_{22}^{S=1} \otimes \mathbb{V}_{11}^{S=0}. \label{eq:fusion-rule-for-D=3}
\end{align}
Evidently, $\mathbb{V}_{\mathbf{T}_3}^{(11)}$ is made of two intermediate spin-1's while two intermediate spin-2's are used to construct $\mathbb{V}_{\mathbf{T}_3}^{(22)}$. Based on the above fusion rules, the site tensors can be built up as

\begin{align}
	&[\mathbf{T}^{(11)}_{3}]_{lrud}^p = C_{1,m_l;1,m_r}^{1,n_1} C_{1,m_u;1,m_d}^{1,n_2} C_{1,n_1;1,n_2}^{1,n_3} C_{1,m_p;1,n_3}^{0,0}   ,\nn
	&[\mathbf{T}^{(22)}_{3}]_{lrud}^p = C_{1,m_l;1,m_r}^{2,n_1} C_{1,m_u;1,m_d}^{2,n_2}  C_{2,n_1;2,n_2}^{1,n_3} C_{1,m_p;1,n_3}^{0,0}  .
	\label{eq:d3_tensor}
\end{align}
The two tensors satisfy

\begin{align}
	&C_4 \circ \mathbf{T}^{(11)}_{3} = +\mathbf{T}^{(11)}_{3},\,\,\,\,\, \sigma \circ \mathbf{T}^{(11)}_{3} = -\mathbf{T}^{(11)}_{3},\nn
	&C_4 \circ \mathbf{T}^{(22)}_{3} = -\mathbf{T}^{(22)}_{3},\,\,\,\,\, \sigma \circ \mathbf{T}^{(22)}_{3} = -\mathbf{T}^{(22)}_{3}.\label{eq:C4-D3}
\end{align} 
All other site tensors, built from fusion rules other than shown in Eq. (\ref{eq:fusion-rule-for-D=3}), are forbidden by $C_4$ and $\sigma$ symmetries. Consequently, with $D=3$, only two distinct classes characterized by quantum numbers, 

\ba \mathbf{T}^{(11)}_{3}: (\theta_{C_4},\theta_{\sigma}, \chi_{\mathcal{T}}, \chi_{\sigma\mathcal{T}}, \chi_{\theta=2\pi}) &=& (+1,-1,+1,+1,+1), \nn
\mathbf{T}^{(22)}_{3}: (\theta_{C_4},\theta_{\sigma}, \chi_{\mathcal{T}}, \chi_{\sigma\mathcal{T}}, \chi_{\theta=2\pi}) &=& (-1,-1,+1,+1,+1), \nonumber \ea
are realizable by means of site tensors $\mathbf{T}^{(11)}_{3}$ and $\mathbf{T}^{(22)}_{3}$, respectively. 

At $D=3$, we discover two distinct classes and within each class, only one possible symmetric tensor. Due to their difference in quantum numbers they cannot be mixed unless one intends to violate the symmetry explicitly. For that matter one cannot mix the $D=3$ tensors with the $D=2$ tensor constructed earlier.

Despite the fact that the two tensor $\mathbf{T}^{(11)}_{3}$ and $\mathbf{T}^{(22)}_{3}$ differ in only one of the quantum numbers, namely $\theta_{C_4}$, the physical properties of the many-body PEPS constructed out of them could not be more different. In an effort to figure out the physical nature of each tensor wave function we have performed finite-size scaling analyses of spin\,($\vec{S}_i$), dimer\,($\delta V_i = \vec{S}_i \cdot \vec{S}_{i+\hat{x}}-\langle \vec{S}_i \cdot \vec{S}_{i+\hat{x}}\rangle $) and vector chirality\,($\delta \vec{\chi}_i = \vec{S}_i \times \vec{S}_{i+\hat{x}}-\langle \vec{S}_i \times \vec{S}_{i+\hat{x}}\rangle $) correlators in finite samples of size $L_x \times L_y$ up to the linear system size $L_x = L_y =48$. Distance between the two operators was fixed at $R=L/3$. Finite MPS-MPO method\,\cite{verstraete04, schollwock11, orus14} with two-site DMRG\,\cite{schollwock11} and zip-up algorithm\,\cite{miles10} were employed to contract the tensor network. The reduced bond dimensions were not fixed in the compression process of MPS, but determined to give an error below $10^{-6}$ after truncated singular value decomposition(tSVD), and three DMRG sweeps are performed. 

The correlators measured in PEPS made of $\mathbf{T}_3^{(11)}$ are presented in Fig.\,\ref{fig:correlator_int_spin1}\,(a)-(c). The perfect exponential decay of all correlators is consistent with the ansatz $\mathbf{T}_3^{(11)}$ representing a fully gapped state devoid of long range order or spontaneous symmetry breaking. We also have evaluated the entanglement entropy to check if topological order is present.  With the cylinder geometry\,(periodic boundary condition along $y$) and using the boundary theory of PEPS\,\cite{cirac11,wahl14}, we find the entanglement entropy obeys the area law with the extrapolation 
$S_{\rm vN}(L_y \rightarrow 0) = 0.06$, in agreement with the absence of topological order. We conclude that PEPS made of $\mathbf{T}_3^{(11)}$ is a featureless, gapped paramagnetic state. The other quantum state realized by $\mathbf{T}_3^{(22)}$ shows totally different aspects in that both spin and dimer correlations fall out algebraically as shown in Fig.\ref{fig:correlator_int_spin2} and only the chirality correlation has exponential decay. We conclude that the PEPS made of $\mathbf{T}_3^{(22)}$ is gapless in both spin singlet and triplet channels.

\section{spin-1 Tensor network states with bond dimension $D=4$ }
\label{sec:d4}

At $D=2$ there was just one symmetric PEPS on a square lattice for spin-1, in agreement with the recent observation~\cite{zaletel16}. At $D=3$, there were two distinct classes of symmetry quantum numbers and only one symmetric site tensor
within each class\,[Eqs. \eqref{eq:d2_sym_tensor} and \eqref{eq:d3_tensor}]. In this section we come to the bond dimension $D=4$, and discover that there can be more classes of symmetric tensors and, more interestingly, can exist a multitude of tensors in a given class. When there are several site tensors sharing the same symmetry quantum numbers, taking their linear combination will not change the symmetry at all but still could result in physically distinct states. In the phrase of Ref.\,\onlinecite{shenghan15A}, there is a family of spin-1 symmetric states sharing the same ``short-distance physics" (dictated by symmetry quantum numbers), yet differing in their long-distance behavior (symmetry breaking or preserving, gapped or critical, etc.)

\subsection{Construction of site tensors}

Let us first show how to construct the site tensors. The bond dimension $D=4$ is achieved either with two flavors of virtual spin-1/2's, or with one flavor of spin-0 and spin-1 each. We focus on the latter case in this article and delegate the first case to a future discussion. The spin-1$\oplus$spin-0 gives a natural extension of the previous construction for $D=3$, which employed a virtual spin-1 only. The bond tensor assumes the block-diagonal form

\begin{align}
	\mathbf{B} =
	\begin{pmatrix}
		1 & 0 & 0 & 0\\ 0 & 0 & 0 & 1 \\
		0& 0 & -1 & 0 \\ 0 & 1 & 0 & 0
	\end{pmatrix} .
\end{align}
The lower $3\times3$ block comes from CG coefficients of mixing two spin-1's into a singlet. Regarding the site tensor, one can follow the procedures used in $D=2,3$ construction to arrive at the following three tensors:

\begin{align}
	&[\mathbf{T}^{(1)}_4]_{lrud}^p = C_{1,m_l;1,m_r}^{2,n_1} C_{1,m_u;1,m_d}^{2,n_2}  C_{2,n_1;2,n_2}^{1,n_3} C_{1,m_p;1,n_3}^{0,0} ,\nn
	&[\mathbf{T}^{(2)}_4]_{lrud}^p = C_{1,m_l;0,m_r}^{1,n_1}C_{0,m_u;1,m_d}^{1,n_2} C_{1,n_1;1,n_2}^{1,n_3}  C_{1,m_p;1,n_3}^{0,0}  ,\nn
	&[\mathbf{T}^{(3)}_4]_{lrud}^p = C_{1,m_l;0,m_r}^{1,n_1} C_{0,m_u;0,m_d}^{0,n_2}  C_{1,n_1;0,n_2}^{1,n_3}  C_{1,m_p;1,n_3}^{0,0} .
	\label{eq:d4_basis}
\end{align}
The fusion processes for each site tensor can be read off from the respective CG coefficients. The first tensor $\mathbf{T}^{(1)}_4$ is the same as the $\mathbf{T}^{(22)}_3$ tensor obtained earlier (note that virtual spin-0 is not used at all in the construction of $\mathbf{T}^{(1)}_4$) and shares the same set of quantum numbers, given by

\ba 
	(\theta_{C_4},\theta_{\sigma}, \chi_{\mathcal{T}}, \chi_{\sigma\mathcal{T}}, \chi_{\theta=2\pi})=(-1,-1,+1,+1,+1).\nn 
	\label{eq:D4-quantum-number}
\ea
Note that $\mathbf{T}^{(2)}_4$ preserves $\sigma$-reflection but breaks $C_4$-rotation, while $\mathbf{T}^{(3)}_4$ breaks both of them. Also, one can show that $\mathbf{T}^{(3)}_4$ satisfies

\begin{align}
	\sigma \circ \mathbf{T}_4^{(3)} = - \mathbf{T}_4^{(3)} = (C_4)^3 \circ \mathbf{T}_4^{(3)} .\nonumber
\end{align}

The broken symmetries in $\mathbf{T}_4^{(2)}$ and $\mathbf{T}_4^{(3)}$ can be ``mended" by taking a linear combination, chosen appropriately so that the final state should have the same $\theta_{C_4} = -1$ and $\theta_{\sigma} = -1$ quantum numbers as $\mathbf{T}^{(1)}_4$. The following combinations

\ba \mathbf{T}^{'(2)}_4  &=& \sum_{n=0}^3 (-C_4)^n \circ \mathbf{T}^{(2)}_4 , \nn
\mathbf{T}^{'(3)}_4  & = & \sum_{n=0}^3 (-C_4)^n \circ \mathbf{T}^{(3)}_4 , \nonumber
\ea
now have all the quantum numbers given in Eq. (\ref{eq:D4-quantum-number}). Furthermore, the three basis tensor just constructed can be arbitrarily combined to produce a continuous family of tensors falling within the same symmetry class (dropping the primes from now on)

\begin{align}
	\mathbf{T}_4 (\bm \lambda) = \cos\! \theta  \mathbf{T}_4^{(1)}
	\!+\! \sin \!\theta  \bigl(\cos \phi \mathbf{T}_4^{(2)} + \sin \phi \mathbf{T}_4^{(3)} \bigr),
	\label{eq:d4_tensor}
\end{align}
where $\bm \lambda = (\theta, \phi)$ symbolizes two parameters of mixing. 

Recall that we assumed a trivial IGG at the outset when trying to classify possible symmetric tensors. As emphasized in Ref.\,\onlinecite{shenghan15B}, the many-body state which results from such classification effort can still possess a non-trivial IGG of emergent character.  Explicitly, one can define a U(1) rotation that gives an arbitrary phase factor $e^{i\phi}$ to virtual spin-1 but  not to virtual spin-0. Under such gauge transformation, $\mathbf{T}_4^{(2)}$ and $\mathbf{T}_4^{(3)}$ are invariant up to overall phase indicating the emergent U(1) IGG. But, one can downgrade or eliminate the U(1) IGG by mixing the basis tensors, and therefore the general tensor in Eq.\,\eqref{eq:d4_tensor} have the trivial IGG for the most part of the parameter space defined by $(\theta,\phi)$. 

\begin{figure}\includegraphics[width=0.45\textwidth]{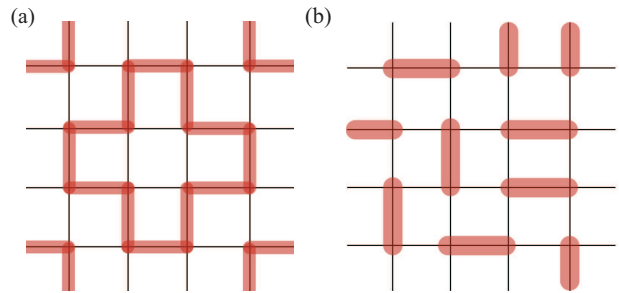}
	\caption{Snapshots of (a) a loop configuration and (b) a dimer configuration that forms the tensor network state formed by tensors $\mathbf{T}_4^{(2)}$ and $\mathbf{T}_4^{(3)}$, respectively. Each red loop in (a) indicates a matrix product state defined in Eq.\,\eqref{eq:t2_loop}, and each red ellipsis in (b) denotes the singlet made out of two spin-1's.}
	\label{fig:t2_state}
\end{figure}

It can be easily shown that the many-body wave function generated by contracting the $\mathbf{T}_4^{(2)}$ tensor is a superposition of loop gas configurations where each loop is a spin-1 chain constructed by projecting two virtual spin-1's into physical spin-1. One of the simplest loop configurations is presented in Fig.\,\ref{fig:t2_state}\,(a). One such closed loop is an MPS state~\cite{zaletel16} 

\begin{align}
	|\phi\rangle = \sum_{\{p_i\}} {\rm Tr} \left[ M_{l_1,r_1}^{p_1} B_{r_1,l_2} M_{l_2,r_2}^{p_2} \cdots  \right]|p_1,p_2,\cdots,p_{N_c}\rangle,
	\label{eq:t2_loop}
\end{align} 
where
\begin{align}
	M_{l_i,r_i}^{p_i} = C_{1,m_{l_i};1,m_{r_i}}^{1,m_{p_i}},\,\,
	B_{r_i,l_j} = C_{1,m_{r_i};1,m_{l_j}}^{0,0},\nonumber
\end{align}
and $N_c$ is the number of sites in a closed loop. Due to the nature of the site tensor $T_4^{(2)}$ there is a constraint in the contraction process of tensors, which forces the loop to turn at 90 degrees at every site. A generic example of a loop obeying the turn-at-every-corner constraint is the cross shown in Fig. \ref{fig:t2_state}(a). Additionally, different loops cannot intersect or touch each other, and still must fill the entire lattice with one loop passing through every site. In the end, there is only one such configuration available, which is formed by forming the smallest-size loop around the $2\times2$ square and tiling them uniformly
over the whole lattice. The resulting plaquette-ordered (PO) state breaks the translation symmetry spontaneously. The full many-body state obtained from the contraction of the $\mathbf{T}_4^{(2)}$ tensor is given, exactly, by 

\begin{align}
	\includegraphics[width=0.48\textwidth]{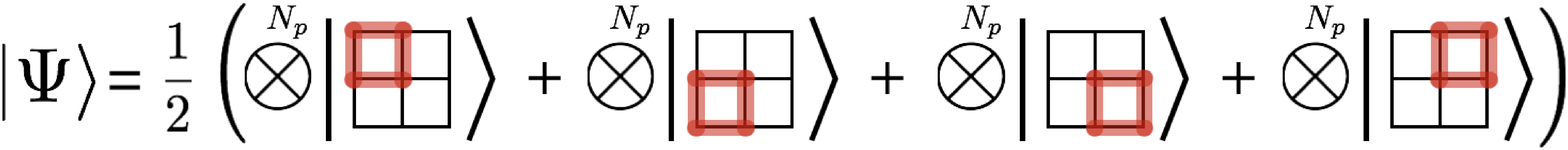} . 
	\label{eq:t2_state}
\end{align}
Each product $\otimes_{N_p}$ is over one-quarter of all the elementary squares in the lattice, and 
\begin{align}
	\includegraphics[width=0.25\textwidth]{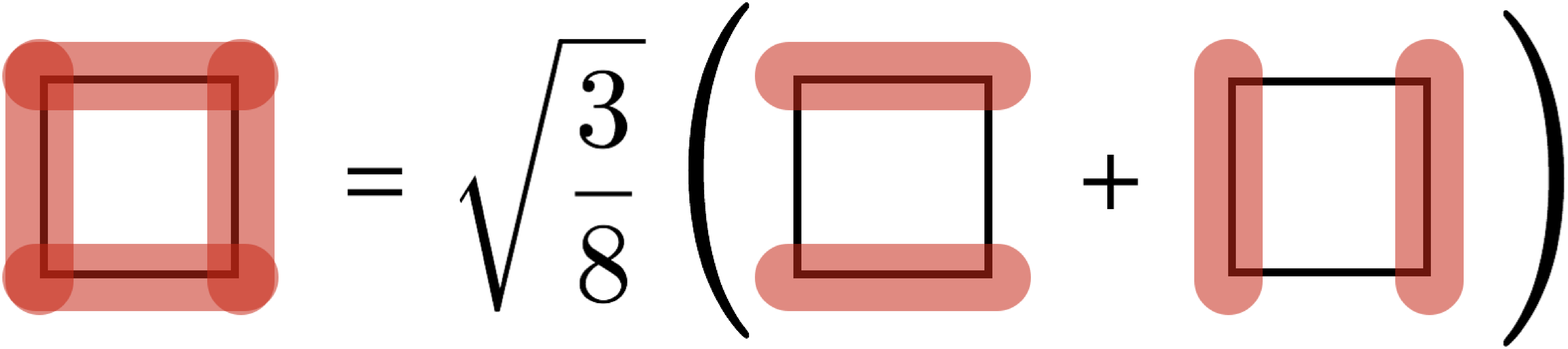} . \nonumber
\end{align}
The positive sign on the right hand side comes from expanding the tensor product explicitly for a small system size. 
Red ellipses denote the spin singlet, $|{\rm singlet}\rangle = (|\uparrow\uparrow\rangle + |\downarrow\downarrow\rangle - |00\rangle)/\sqrt{3}$, and $N_p$ is the number of non-overlapping $2\times2$ plaquettes. More precisely, the state is a Schrodinger's cat-like superposition of four distinct macroscopic states, each of which represents a PO state. The appearance of the Schr\"{o}dinger's cat state is a consequence of the symmetry we insisted upon the tensor. Taking the linear combination restores the translation symmetry violated by an individual PO state. One can evaluate the nearest-neighbor bond strength exactly in terms of the state in Eq.\,\eqref{eq:t2_state},

\begin{align}
	\langle \vec{S}_i \cdot \vec{S}_{i+\hat{x}\,(\hat{y})} \rangle = -5/8 .\nonumber	
\end{align}
The same number has been found by numerical contraction on (odd)$\times$(odd) sites\,(up to $51 \times 51$) of the tensor $\mathbf{T}_4^{(2)}$. On the other hand, the (even)$\times$(even) lattice allows only one of the configurations in Eq. (\ref{eq:t2_state})
to materialize, and the bond strength changes suddenly to 

\begin{align}
	\langle \vec{S}_i \cdot \vec{S}_{i+\hat{x}\,(\hat{y})} \rangle = -5/4 .\nonumber	
\end{align}
Indeed this is what we get with numerical calculation on the (even)$\times$(even) lattice.

The many-body state produced by $\mathbf{T}_4^{(3)}$ is the nearest neighbor resonating valence bond state\,(NN RVB). Since the physical spin forms the singlet with a virtual spin-1 on one of virtual legs while enforcing other legs to have spin-0\,[Eq.\,\eqref{eq:d4_basis}],  the contraction of neighboring $\mathbf{T}_4^{(3)}$'s gives rise to the nearest neighbor singlet of physical spin-1's or dimer configuration as depicted in Fig.\,\ref{fig:t2_state}\,(b). Therefore, the resulting PEPS after the contraction over the whole lattice is a superposition of all possible dimer configurations, or NN RVB. By numerical analyses, we found that in the state made of $\mathbf{T}_4^{(3)}$ the spin correlation decays exponentially with the finite correlation length $\xi=0.70$, while the dimer correlation decay algebraically with the exponent $\alpha = 1.56$. These are in excellent agreement with the ones observed by quantum Monte Carlo method for the same NN RVB state in Ref.\,\onlinecite{melko13}.

\subsection{``Phase diagram"}

Now that we identified the quantum states built with each basis tensor $\mathbf{T}_4^{(i)}$\,($i=1,2,3$), let us turn to general states realized by $\mathbf{T}_4 (\bm \lambda)$ in Eq. (\ref{eq:d4_tensor}). There is a notion of ``phase diagram" one can introduce in this family of states, treating $(\theta, \phi)$ as the tuning parameters. In principle, a phase boundary separating distinct phases can exist in such putative phase diagram~\cite{shenghan15A}. 

An efficient way to carve out the phase boundary within this family of tensors is to use the fidelity measure\,\cite{zanardi06, zhou08a, zhou08b}, which is the overlap of two ground state wave functions separated slightly in the parameter space. The fidelity is known to exhibit a significant drop or a non-analytic behavior when the two states belong to different sides of the phase boundary, or on the same side but sufficiently close to such a boundary. The numerical cost of identifying the phase boundary through the fidelity calculation is a lot less expensive than the conventional one involving the finite-size scaling of various correlators\,\cite{zanardi06, ning08, zhou08a, zhou08b, gu10}. Once all the phase boundaries are safely identified through the fidelity calculation, it will be sufficient to do the analyses of order parameters and correlation functions only on a select set of points to pin down the nature of a given phase. 

\begin{figure}\includegraphics[width=0.5\textwidth]{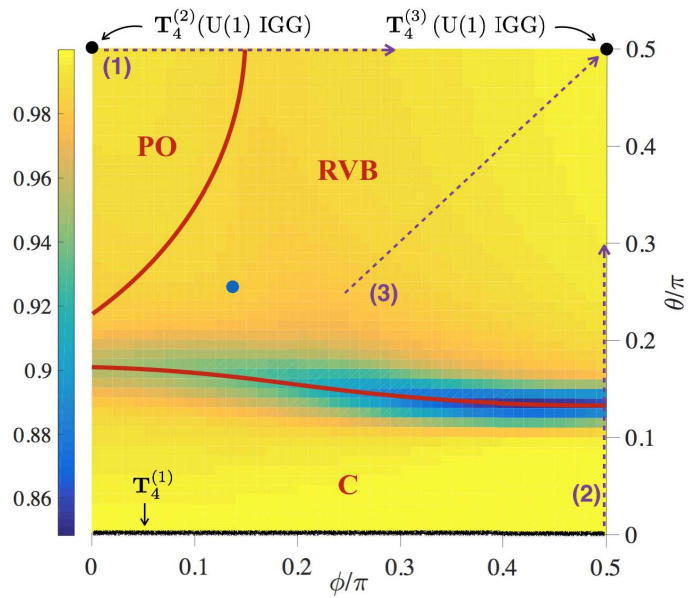}
	\caption{Quantum phase diagram of the class characterized by $(\theta_{C_4},\theta_{\sigma}, \chi_{\mathcal{T}}, \chi_{\sigma\mathcal{T}}, \chi_{\theta=2\pi})=(-1,-1,+1,+1,+1)$ as a function of two tunable parameters $\theta$ and $\phi$\,[Eq.\,\eqref{eq:d4_tensor}].  On the $\theta=0$ line, the PEPS is made of solely $\mathbf{T}_4^{(1)}$ while $\mathbf{T}_4^{(2)}$ and $\mathbf{T}_4^{(3)}$ at the point $(\theta,\phi)=(\pi/2,0)$ and $(\pi/2, \pi/2)$, respectively. Here, the blue dot at $(\theta,\phi) = (0.26\pi,0.15\pi)$ denotes the energetically most favorable state for $J_1-J_2$ Hamiltonian with $J_2=0.54J_1$ and its energy density is $E = -1.44J_1$. Along the dotted purple lines specified by (1)-(3), various correlations and physical quantities are measured to identify the quantum phases, and results are shown in Fig.\,\ref{fig:po2rvb2critical}.}
	\label{fig:phase_diagram}
\end{figure}
\begin{center}
  \begin{table}
    \begin{tabular}{| c || c | c |}
      \hline
      phase & spin & dimer \\
      \hline \hline
      C & pow & pow \\ \hline
      PO & exp & exp  \\ \hline
      Gapped RVB & exp & exp \\ \hline
      Critical RVB ($\mathbf{T}_4^{(3)})$ & exp & pow \\ \hline
      \end{tabular}
    \caption{Long-range behavior of correlation functions in each phase, where pow and exp denote power-law and exponentially decaying functions, respectively. Vector spin chirality correlation is exponential for all the phases above. }
\label{table:correlation}
  \end{table}
\end{center}

So far, we have not specified any microscopic Hamiltonian tied to the specific tensor $\mathbf{T}_4 (\bm \lambda)$. It has been claimed that a parent Hamiltonian having a given tensor network state as its ground state always exists\,\cite{perez08,swingle10,schuch10}. Therefore, one can view the phase diagram we are constructing as that of some putative Hamiltonian $H(\bm \lambda)$, whose ground state is the one given by the contraction of $\mathbf{T}_4 (\bm \lambda)$.  With this philosophical backing, we calculate the {\it fidelity per lattice site}\,\cite{zanardi06, zhou08a, zhou08b}

\begin{align}
	f(\bm \lambda) = \lim_{N\rightarrow \infty} f_N (\bm \lambda) = \lim_{N\rightarrow \infty} \exp \left[ \frac{\ln F_N(\bm \lambda, \bm \lambda + \delta\bm \lambda)}{N} \right],
	\label{eq:fidelity}
\end{align}
where $N=L_x \times L_y$ is the number of lattice sites, $\delta \bm \lambda = (\delta \theta, \delta \phi)$ is a small deviation from $\bm \lambda$, and 

\ba F_N (\bm \lambda, \bm \lambda +\delta \bm \lambda) = | \langle\psi(\bm \lambda) | \psi(\bm \lambda +\delta \bm \lambda)\rangle | \nonumber \ea 
is the fidelity obtained from the overlap of two PEPS states $|\psi(\bm \lambda)\rangle$ made out of 
$\mathbf{T}_4 (\bm \lambda)$. The SU(2)-invariant tensor algorithm established in Ref.\,\cite{vidal12} was used to contract the tensor network of $\langle\psi(\bm \lambda) | \psi(\bm \lambda +\delta \bm \lambda)\rangle$ with fixed $\delta \bm \lambda = (0.01,0.01)$, and we have employed the same error threshold as the one adopted in the previous section. 

Fidelity calculation clearly shows the signature of phase transitions or phase boundaries; see Fig. \ref{fig:phase_diagram}. Further calculations of the various correlation functions allow us to identify each phase as C, RVB, and PO. In the C (critical) phase both spin and dimer correlators decay algebraically, while in the RVB phase both correlators are exponentially decaying. The region in the vicinity of the $\mathbf{T}_4^{(2)}$ state is found to be the plaquette-ordered phase. 


%
\begin{figure}	\includegraphics[width=0.5\textwidth]{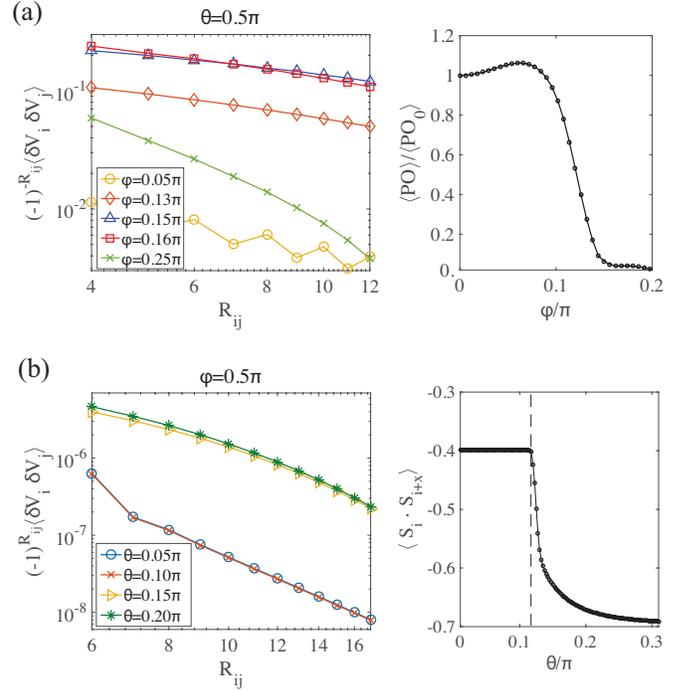}
	\caption{Dimer correlation functions for several (a) $\phi$'s at $\theta=0.5\pi$\,[(1)-line in Fig.\, \ref{fig:phase_diagram}] and (b) $\theta$'s at $\phi=0.5\pi$\,[(2)-line in Fig.\, \ref{fig:phase_diagram}]. Right panels of (a) and (b) are the plaquette order parameter\,[see Eq.\,\eqref{eq:po_order}] and the energy density for Heisenberg exchange interaction, respectively. Here, $\langle {\rm PO}_0\rangle=-5/4$ denotes the expectation of the plaquette order parameter for the state made of $\mathbf{T}_4^{(2)}$, i.e. $(\theta,\phi)=(0.5\pi,0)$. As for the correlations, the system size is chosen to be $L_x=L_y=3R_{ij}$ for each $R_{ij}$, while $L_x=L_y=40$ is fixed for measuring the plaquette order parameter and energy density. }
	\label{fig:po2rvb2critical}
\end{figure}

In addition to the measurement of spin, dimer, and the vector chirality correlators as in the previous sections, we have introduced an additional, plaquette order parameter defined by

\begin{align}
	{\rm PO}_i & = \frac{1}{2} \Big(\vec{S}_{i}\cdot \vec{S}_{i+\hat{x}} - \vec{S}_{i+\hat{x}}\cdot \vec{S}_{i+2\hat{x}}\nn
	&~~~~~~
	+\vec{S}_{i}\cdot \vec{S}_{i+\hat{y}} - \vec{S}_{i+\hat{y}}\cdot \vec{S}_{i+2\hat{y}}\Big),
	\label{eq:po_order}
\end{align}
which measures the bond modulation strength in the $x$- and $y$-directions simultaneously.  
In both PO and RVB phases,  spin and vector chirality correlators decay exponentially. However, the correlation length of dimer correlator diverges as the PO/RVB phase boundary is approached, as shown in Fig.\,\ref{fig:po2rvb2critical}\,(a). The dimer correlators are measured along the line (1) in Fig.\,\ref{fig:phase_diagram}. Calculation of the PO parameter on a $40\times40$ lattice is shown in Fig.\,\ref{fig:po2rvb2critical}\,(a), displaying a continuously vanishing order parameter around the critical value $\phi_c \simeq 0.15\pi$. Calculations at other values of $\theta < \pi/2$ showed an entirely similar results as the one in Fig.\,\ref{fig:phase_diagram}\,(a). Overall we conclude that there is a continuous phase transition between PO and RVB phases in the phase diagram of $\mathbf{T}_4 (\bm \lambda)$. 

As for the RVB and C phase boundaries, all numerical signatures point to the first order phase transition. As shown in Fig.\,\ref{fig:po2rvb2critical}\,(b), the behavior of dimer correlations changes abruptly from algebraic (C) to exponentially decaying (RVB) on crossing the phase boundary along the line (2) in Fig. \ref{fig:phase_diagram}. Measurement of the bond strength $\langle \vec{S}_i \cdot \vec{S}_{i+\hat{x}}\rangle$ also shows a non-analytic behavior  at the critical point $(\theta,\phi)=(0.12\pi,0.5\pi)$ as shown on the right of Fig.\,\ref{fig:po2rvb2critical}\,(b). Spin correlations show similar abrupt changes, but the vector chirality correlations decay exponentially on both sides. Calculations across the C/RVB boundaries elsewhere showed similar abrupt changes. 

\begin{figure}	\includegraphics[width=0.35\textwidth]{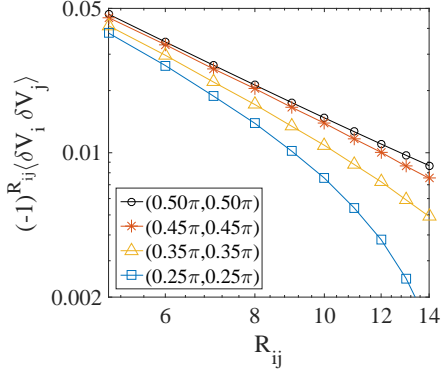}
	\caption{Dimer correlation functions along the line (3) in Fig.\,\ref{fig:phase_diagram}, where the legend specifies the parameters $(\theta,\phi)$. The linear system size was chosen to be $L_x=L_y=3R_{ij}$.}
	\label{fig:gap_rvb2crit_rvb}
\end{figure}

Finally, as mentioned earlier, $\mathbf{T}_4^{(3)}$ gives the NN RVB state that shows an algebraically decaying dimer correlation and exponentially decaying spin correlation. The criticality of the dimer correlation is unique to the $\mathbf{T}_4^{(3)}$, while the rest of the RVB region has exponentially decaying dimer correlation. In support of this claim we present in Fig.\,\ref{fig:gap_rvb2crit_rvb} how the dimer correlations evolve in approaching the critical RVB state from the gapped RVB state along the dashed line (3) in Fig.\,\ref{fig:phase_diagram}. Note that the correlation length of dimer correlator increases rapidly in approaching the NN RVB state.      

Spin liquid phase of the spin-1 state is most often discussed in the context of  the $J_1\! -\! J_2$ Hamiltonian,

\begin{align}
	H = J_1 \sum_{\langle i,j\rangle} \vec{S}_i\cdot\vec{S}_j 
	+ J_2 \sum_{\langle\!\langle i,j \rangle\!\rangle} \vec{S}_i\cdot\vec{S}_j, \label{eq:J1-J2-H}
\end{align}
with $J_1 >0 $ and $J_2 >0 $ for the nearest and the diagonal neighbor interactions, respectively. Earlier DMRG work of Jiang et al. found a paramagnetic phase around $J_2\simeq 0.54J_1$ and the energy per site $E\simeq - 1.46J_1$\,\cite{jiang09}. We are curious to see if the spin-1 phase diagram we constructed contains states that have competitive energies. To test this out we have employed the {\it simplex} algorithm to find an energy minimum point in our parameter space for the $J_1-J_2$ model on a $60\times60$ lattice with $J_2=0.54J_1$. Energy-optimizing point was found at $(\theta,\phi)=(0.26\pi,0.15\pi)$, where the energy per site was $E=-1.44J_1$, in close competition to the DMRG value. Correlation function calculations were not carried out in the DMRG work due to the limited system size. Our result for the correlation function at the energy-optimizing point indicates that the fully symmetric paramagnet state could be a candidate ground state to the nematic paramagnet found in the DMRG study. 

\subsection{Other states}

Construction of the $D=4$ symmetric spin-1 state has so far been limited to the sub-Hilbert space in which the quantum numbers characterizing the states were $(\theta_{C_4},\theta_{\sigma}, \chi_{\mathcal{T}}, \chi_{\sigma\mathcal{T}}, \chi_{\theta=2\pi})=(-1,-1,+1,+1,+1)$. Carrying out a similar analysis as before for another symmetry class, in which $(\theta_{C_4},\theta_{\sigma}, \chi_{\mathcal{T}}, \chi_{\sigma\mathcal{T}}, \chi_{\theta=2\pi}) = (+1,-1,+1,+1,+1)$, gives the following three basis tensors

\begin{align}
	&[\mathbb{T}^{(1)}_4]_{lrud}^p = C_{1,m_l;1,m_r}^{1,n_1} C_{1,m_u;1,m_d}^{1,n_2}  C_{1,n_1;1,n_2}^{1,n_3} C_{1,m_p;1,n_3}^{0,0} ,\nn
	&[\mathbb{T}^{(2)}_4]_{lrud}^p = C_{1,m_l;0,m_r}^{1,n_1}C_{0,m_u;1,m_d}^{1,n_2} C_{1,n_1;1,n_2}^{1,n_3}  C_{1,m_p;1,n_3}^{0,0}  ,\nn
	&[\mathbb{T}^{(3)}_4]_{lrud}^p = C_{1,m_l;1,m_r}^{1,n_1} C_{0,m_u;1,m_d}^{1,n_2}  C_{1,n_1;1,n_2}^{1,n_3}  C_{1,m_p;1,n_3}^{0,0},
	\nonumber
\end{align}
where $\mathbb{T}_4^{(1)}=\mathbf{T}_3^{(11)}$, $\mathbb{T}_4^{(2)}=\mathbf{T}_4^{(2)}$ and $\mathbb{T}_4^{(3)}$ is a new tensor which can survive after symmetrization to have $(\theta_{C_4},\theta_{\sigma})=(+1,-1)$. Therefore, a general symmetric tensor is given by

\begin{align} 
	\mathbb{T}_4 = \mathbb{T}_4^{(1)} + \sum_{n=0}^3 (C_4)^n \circ \left( \alpha\mathbb{T}_4^{(2)} + \beta \mathbb{T}_4^{(3)} \right),
	\label{eq:second-D4-tensor}
\end{align}
where $\alpha$ and $\beta$ are arbitrary real numbers. This is the extension of the gapped state built by $\mathbf{T}_3^{(11)}$ since they share the same quantum number. One can show that $\mathbb{T}^{(2)}$ and $\mathbb{T}^{(3)}$ both have the emergent U(1) IGG. Numerics of fidelity and correlation function calculations on a family of states (\ref{eq:second-D4-tensor}) might yield another sort of interesting phase diagram. 

We can also construct the symmetric PEPS's with two flavor of spin-1/2's as mentioned earlier. Those states should carry the quantum number $\theta_{\theta=2\pi}=-1$ which can never be realized with integer virtual spins. It cannot be confirmed before investigation, but there might be a possibility to finding a better ansatz for $J_1-J_2$ Hamiltonian than the one found above. We may also expect to discovering some exotic phases. But, we leave those interesting subjects for the future work. 

\section{Summary and Discussion}
A systematic classification of symmetric tensor network states respecting all the lattice symmetries of the square lattice and two internal symmetries, time reversal and spin rotation, has been carried out in this paper. We found that $2^5$ distinct classes can exist, each of them characterized by five binary quantum numbers $(\theta_{C_4},\theta_{\sigma}, \chi_{\mathcal{T}}, \chi_{\sigma\mathcal{T}}, \chi_{\theta=2\pi})$. Also, essential constraints\,[Eq.\,\eqref{eq:constraints}] on virtual Hilbert space were found to build symmetric states. It was also shown that the integer and half-integer spins cannot be accommodated on the virtual legs simultaneously as long as the trivial IGG is imposed. 

We derived exemplary symmetric tensor states for the bond dimensions $D=2,3,4$. First, it was proven that only one class characterized by $(\theta_{C_4},\theta_{\sigma}, \chi_{\mathcal{T}}, \chi_{\sigma\mathcal{T}}, \chi_{\theta=2\pi})=(-1,-1,+1,+1,-1)$ can be allowed with $D=2$. This class is represented by a single basis tensor $\mathbf{T}_2^{(1,1)}$, which is identical to the ansatz proposed as a featureless spin-1 quantum paramagnet in Ref.\,\onlinecite{zaletel16}.

With $D=3$, we are able to construct two distinct classes characterized by $(\theta_{C_4},\theta_{\sigma}, \chi_{\mathcal{T}}, \chi_{\sigma\mathcal{T}}, \chi_{\theta=2\pi}) = (+1,-1,+1,+1,+1)$ and $(-1,-1,+1,+1,+1)$. By calculating the spin, dimer and vector chirality correlations, we were able to show that the $\theta_{C_4}=+1$ state has all the correlations decaying exponentially, and have the topological entanglement entropy nearly zero. This is a fully gapped liquid state without topological order. The other state with $\theta_{C_4}=-1$ has critical correlations in both spin and dimer channels. 
\begin{center}
  \begin{table}
    \begin{tabular}{| c | c |c || c | c |}
      \hline
      D & $(\theta_{C_4},\theta_{\sigma}, \chi_{\mathcal{T}}, \chi_{\sigma\mathcal{T}}, \chi_{\theta=2\pi})$ & state & $\langle \vec{S}_i \cdot \vec{S}_{i\!+\!\hat{x}}\rangle$ & $\langle \vec{S}_i \cdot \vec{S}_{i\!+\!\hat{x}\!+\!\hat{y}}\rangle$  \\
      \hline \hline
       2 & (-1,-1,+1,+1,-1) & gapped &  -0.4620 & -0.1740 \\ \hline
      3 & (+1,-1,+1,+1,+1) & gapped & -0.1545 & -0.0313 \\ \hline
      3 & (-1,-1,+1,+1,+1) & gapless & -0.3969 & -0.1555  \\ \hline
      4 & (-1,-1,+1,+1,+1) & gapped & -0.9105 & +0.3545  \\ \hline
      \end{tabular}
    \caption{Nearest-neighbor and diagonal bond strength averages for the spin-1 PEPS. Numbers quoted for $D=4$ are from the optimized wave function at $(\theta, \phi) = (0.26\pi, 0.15\pi)$.}
\label{table:energy_density}
  \end{table}
\end{center}

Up to $D=3$, only a single basis tensor could exist for each class, representing a unique symmetric quantum state for that set of symmetry quantum numbers. In the $D=4$ construction, we were able to find three independent basis tensors in the same class with $(\theta_{C_4},\theta_{\sigma}, \chi_{\mathcal{T}}, \chi_{\sigma\mathcal{T}}, \chi_{\theta=2\pi}) = (-1,-1,+1,+1,+1)$. Other symmetry class showed a similar potential to host several tensor states. Our findings imply that by increasing the bond dimension one could construct a multitude of quantum states sharing the same short range physics. By taking linear combinations of the basis states and studying their long-range behavior, i.e. correlation functions, in the space of mixing parameters, one could construct a ``phase diagram" hosting a number of distinct phases separated by quantum phase transitions. We were able to carry out such a program explicitly through the adoption of the fidelity idea. The phase diagram we worked out in the $(\theta_{C_4},\theta_{\sigma}, \chi_{\mathcal{T}}, \chi_{\sigma\mathcal{T}}, \chi_{\theta=2\pi}) = (-1,-1,+1,+1,+1)$ space hosts three distinct phases: plaquette-ordered, gapped RVB, and critical phase. Nature of the phase diagram among the phases can be worked out employing the conventional analysis of correlation functions. Despite the significant amount of numerical effort we invested in working out the phase diagram, there remains a challenging task of studying possible phase diagrams and critical behavior in other symmetry sectors, which we invite the readers to share. 

The transition between the plaquette-ordered and the gapped RVB phase is second-order according to our careful numerical study. 
An earlier PEPS work looking at the phase transition between the Neel-ordered phase and the valence bond phase indicated the possibility of a deconfined quantum critical point between them\,\cite{nina_wang11}. At this point, we are not sure that the deconfined transition scenario applies to the PO-RVB transition in our model study. We do not have any reason to believe that the RVB state will eventually Neel-order at distances much longer than we were able to examine, and indeed all spin-spin correlation functions were consistent with the exponential decay. A transition out of the plaquette-ordered phase may be driven by the proliferation of topological defects, if one assumes that the scenario worked out for the melting of valence bond order\,\cite{levin04} should apply to the melting of the plaquette order as well. 
At this point, we are not certain that a well-defined field-theoretic description of the PO-RVB phase transition is available. If so, it will be an exciting problem for the near future to work out. 

A potential use of the various wave functions we constructed is as variational wave functions of frustrated spin Hamiltonian on a square lattice. As mentioned in the previous section, the $J_1-J_2$ spin Hamiltonian (\ref{eq:J1-J2-H}) is believed to host QSL phase near $J_2 \approx J_1 /2$. The nearest and the diagonal bond strengths calculated for each of the tensor states we constructed are listed in Table\,\ref{table:energy_density}. It was shown that a suitable mixture of $D=4$ tensors can generate a state with very competitive energy as the one found in the DMRG study. 

\acknowledgements{The authors are indebted to Shenghan Jian and Ying Ran for helpful discussions. Hyunyong Lee is supported by the NRF grant (No.2015R1D1A1A01059296)}

\appendix

\section{Classification on square lattice}
\label{appendix:classification}

We start with considering a {\it symmetric} PEPS where site and bond tensors satisfy\,\cite{swingle10, shenghan15A}

\begin{align}
	\mathbf{T} &= \Theta_R W_R R \circ \mathbf{T},\nn
	\mathbf{B} &= W_R R \circ \mathbf{B},
	\label{eq:sym_peps}
\end{align}
where $\mathbf{T}$ and $\mathbf{B}$ respectively denote site and bond tensors, $W_R$ and $\Theta$ respectively indicate the gauge transformation and overall U(1) phase factor associated with symmetry operation $R$\,\cite{shenghan15A,shenghan15B}.  

Spatial symmetry of square lattice is fully defined by 4 generators: $\{T_1,\,T_2, C_4, \sigma \} $ corresponding translation along $x$- and $y$-directions, $\pi/2$-rotation around the origin site and reflection along $y=x$ axis such that 

\begin{align}
	T_1 : (x,y,i) &\rightarrow (x+1,y,i),\nn
	T_2 : (x,y,i) &\rightarrow (x,y+1,i),\nn
	C_4 : (x,y,i) &\rightarrow (-y,x,i_{C_4}),\nn
\sigma : (x,y,i) &\rightarrow (y,x,i_{\sigma}),\nn
\end{align}
where $(x,y,i)$ denotes the leg index $i=\{l,r,u,d\}$ of site tensor at position $(x,y)$ and

\begin{align}
	\{l_{C_4},r_{C_4},u_{C_4},d_{C_4} \} &= \{ d, u, l, r \},\nn
	\{l_{\sigma},r_{\sigma},u_{\sigma},d_{\sigma} \} &= \{ d, u, r, l \}
\end{align}
Fig.\,\ref{fig:schematic} schematically depicts the action of each generator on square lattice and site tensor. These four generators completely define space group by the following commutative relations

\begin{align}
	T_2^{-1} T_1^{-1} T_2 T_1 &= e,\nn
	C_4^{-1} T_1 C_4 T_2 &= e,\nn
	C_4^{-1} T_2 C_4 T_1^{-1} &= e,\nn
	(C_4)^4 &= e,\nn
	\sigma^{-1} T_2^{-1} \sigma T_1 &= e,\nn
	\sigma^{-1} T_1^{-1} \sigma T_2 &= e,\nn
	(\sigma)^2 &=e,\nn
	\sigma^{-1} C_4 \sigma C_4 &= e.
	\label{eq:space_groupA}
\end{align}
Since the time reversal\,($\mathcal{T}$) and SU(2) spin rotation\,($U_{\theta\vec{n}}$) commute with each other as well as all space symmetry operations, one can consider the following relations

\begin{align}
	g^{-1} \mathcal{T}^{-1} g \mathcal{T} &= e,\,\,\forall g=T_1,T_2,C_4,\sigma,\nn
	g^{-1} U_{\theta\vec{n}}^{-1} g U_{\theta\vec{n}} &= e,\,\,\forall g=T_1,T_2,C_4,\sigma,\mathcal{T}.
	\label{eq:time_spin_rotationA}
\end{align}
These commutative relations provide algebraic equations of $W_R$ and $\Theta_R$ associated with symmetry operation $R=\{T_1,T_2,C_4,\sigma,\mathcal{T},U_{\theta\vec{n}} \}$\,\cite{shenghan15A}. By fixing gauge, one can choose specific form of each gauge transformations satisfying all the algebraic equations.

Let us consider the transformation rules generated by the space group generators in Eq.\,\eqref{eq:space_group}. The relation $T_2^{-1} T_1^{-1} T_2 T_1 = e$ gives the following algebraic equation\,\cite{shenghan15A, shenghan15B}

\begin{align}
	&W_{T_2}^{-1}(T_2(x,y,i)) W_{T_1}^{-1} (T_1T_2(x,y,i)) \nn
	&\times W_{T_2}(T_1T_2(x,y,i)) W_{T_1} (T_1(x,y,i)) = \chi_{12}(x,y,i),\nn 
	&\Theta_{T_2}^*(T_2(x,y)) \Theta_{T_1}^* (T_1T_2(x,y)) \nn
	&\times \Theta_{T_2}(T_1T_2(x,y)) \Theta_{T_1} (T_1(x,y)) = \prod_i \chi_{12}^*(x,y,i).
	\label{eq:t12_relation}
\end{align}
First, the gauge redundancy or $V$- and $\Phi$-ambiguities\,\cite{shenghan15A} can be used to fix $W_{T_2}(x,y,i)$ and $\Theta_{T_2}(x,y)$ as follows

\begin{align}
	W_{T_2}(x,y,i) &\rightarrow V(x,y,i) W_{T_2}(x,y,i) V^{-1}(T_2(x,y,i)),\nn
	\Theta_{T_2}(x,y) &\rightarrow \Phi(x,y) \Theta_{T_2}(x,y) \Phi^*(T_2(x,y)),
	\label{eq:fixing_wt2}
\end{align}
and we choose $W_{T_2}(x,y,i) = \mathbb{I}$ and $\Theta_{T_2}(x,y) = 1$. Then, in this gauge, it is easy to prove using Eq.\,\eqref{eq:sym_peps} that the site tensor does not depend on $y$ coordinate, i.e. $\mathbf{T}(x,y,i) = \mathbf{T}(x,0,i)$. We can infer from the above gauge transformation that the remaining $V$- and $\Phi$-ambiguities maintaining $y$-independent tensor network or $W_{T_2} = \mathbb{I}$ and $\Theta_{T_2} = 1$ should satisfy

\begin{align}
	&V(x,y,i) = V(x,y+1,i) = V(x,0,i),\nn
	&\Phi(x,y) = \Phi(x,y+1) = \Phi(x,0),
	\label{eq:v_t2}
\end{align}
Furthermore, $\varepsilon_{T_{1,2}}$-ambiguity, i.e. $W_{T_{1,2}} \rightarrow \varepsilon_{T_{1,2}} W_{T_{1,2}}$, allows us to choose $\chi_{12}(x,y,i)$ as we like such that

\begin{align}
	\chi_{12}(x,y,i) &\rightarrow \varepsilon_{T_2}^*(T_2(x,y,i)) \varepsilon_{T_1}^*(T_1 T_2(x,y,i)) \nn	& \times\varepsilon_{T_2}(T_1T_2(x,y,i)) \varepsilon_{T_1}(T_1(x,y,i))\,\chi_{12}(x,y,i).
	\label{eq:fixing_chi12}
\end{align}
For simplicity, we fix $\chi_{12}(x,y,i)=1$, and then Eq.\,\eqref{eq:t12_relation} can be recast as

\begin{align}
	&W_{T_1}(x\!+\!1,y\!+\!1,i) = W_{T_1} (x\!+\!1,y,i),\nn 
	&\Theta_{T_1}(x\!+\!1,y\!+\!1) = \Theta_{T_1} (x\!+\!1,y),
\end{align}
indicating $W_{T_1}(x,y,i)=W_{T_1}(x,0,i)$ and $\Theta_{T_1}(x,y)=\Theta_{T_1}(x,0)$. Since the remaining $V$- and $\Phi$-ambiguities are also $y$-independent, we can use them to fix $W_{T_1}(x,y,i)$ and $\Theta_{T_1}(x,y)$ to be identity and unity, respectively, and therefore

\begin{align}
	W_{T_{1,2}}(x,y,i) = \mathbb{I},\;\;\;\;\;\;\; \Theta_{T_{1,2}} = 1.
		\label{eq:sol_t12}
\end{align} 

Now, Eq.\,\eqref{eq:sym_peps} for $T_1$ tells us that in this gauge the site tensor is site-independent: $\mathbf{T}(x,y,i)=\mathbf{T}(i)$. One can also easily prove that the remaining $V$- and $\Phi$-ambiguities keeping the site-independence of PEPS\,($W_{T_{1,2}}=\mathbb{I}$ and $\Theta_{T_{1,2}}=1$) should be also site-independent: $V(x,y,i) = V(i)$ and $\Phi(x,y)=\Phi$.

Considering $C_4^{-1} T_1 C_4 T_2 = e$ and Eq.\,\eqref{eq:sol_t12}, we would obtain

\begin{align}
	&W_{C_4}^{-1}(x,y,i) W_{C_4}(T_1^{-1}(x,y,i)) = \chi_{T C_4}^{(1)}(C_4^{-1}(x,y,i)),\nn 
	&\Theta_{C_4}^*(x,y) \Theta_{C_4}(T_1^{-1}(x,y)) = \prod_i [\chi_{TC_4}^{(1)}(C_4^{-1}(x,y,i))]^*.
	\label{eq:tc4_relation1}
\end{align}
%
As done in Eq.\,\eqref{eq:fixing_chi12}, we can fix $\chi_{TC_4}^{(1)}(x,y,i)=1$ using the $\varepsilon_{C_4}$-ambiguity\,($W_{C_4} \rightarrow \varepsilon_{C_4} W_{C_4}$), and then find from Eq.\,\eqref{eq:tc4_relation1} that

\begin{align}
	&W_{C_4}(x,y,i) = W_{C_4}(x\!-\!1,y,i)= W_{C_4}(0,y,i),\nn
	&\Theta_{C_4}(x,y) = \Theta_{C_4}(x\!-\!1,y) = \Theta_{C_4}(0,y).
	\label{eq:wc4_x}
\end{align}
Now, remaining $\varepsilon_{C_4}$-ambiguity preserving $\chi_{TC_4}^{(1)}(x,y,i)=1$ should satisfy

\begin{align}
	\varepsilon_{C_4}(x,y,i) = \varepsilon_{C_4}(x\!-\!1,y,i) = \varepsilon_{C_4}(0,y,i).
	\label{eq:ec4_x}
\end{align}
Similarly, inserting Eqs.\,\eqref{eq:sol_t12} and \eqref{eq:wc4_x} into an algebraic equation from $C_4^{-1} T_2 C_4  T_1^{-1} = e$, we can find

\begin{align}
	&W_{C_4}^{-1}(0,y,i) W_{C_4}(0,y\!-\!1,i) = \chi_{TC_4}^{(2)} (C_4^{-1}(x,y,i)),\nn
	&\Theta_{C_4}^*(0,y) \Theta_{C_4}(0,y\!-\!1) = \prod_i [\chi_{TC_4}^{(2)} (C_4^{-1}(x,y,i))]^*.
	\label{eq:tc4_relation2}
\end{align}
Since LHS of the above equation does not depend on $x$-coordinate, $\chi_{TC_4}^{(2)}(C_4^{-1}(x,y,i)) = \chi_{TC_4}^{(2)}(C_4^{-1}(0,y,i))$ and therefore we can use the remaining $\varepsilon_{C_4}$-ambiguity\,[Eq.\,\eqref{eq:ec4_x}] to fix $\chi_{TC_4}^{(2)}(C_4^{-1}(0,y,i)) = 1$. Now, Eq.\,\eqref{eq:tc4_relation2} read

\begin{align}
	&W_{C_4}(0,y,i) = W_{C_4}(0,y\!-\!1,i) = W_{C_4}(0,0,i) \equiv w_{C_4}(i),\nn
	&\Theta_{C_4}(0,y) = \Theta_{C_4}(0,y\!-\!1) = \Theta_{C_4}(0,0) \equiv \theta_{C_4}(i).
\end{align}
Inserting Eq.\,\eqref{eq:wc4_x} into the above equation, we obtain

\begin{align}
	W_{C_4}(x,y,i) = w_{C_4}(i),\;\;\;\;\;\;\; \Theta_{C_4}(x,y) = \theta_{C_4}(i),
\end{align}
leading the following equations from the group relation $(C_4)^4=e$ that

\begin{align}
	&w_{C_4}(l) w_{C_4}(u) w_{C_4}(r) w_{C_4}(d) = \chi_{C_4}(i),\nn
	&(\theta_{C_4})^4 = \prod_i \chi_{C_4}^*(i).
	\label{eq:wc4_relation}
\end{align}
Since LHS of the fist equation in Eq.\,\eqref{eq:wc4_relation} is independent on leg index\,$i$, $\chi_{C_4}(i) = \chi_{C_4}$ is also leg independent implying $\chi_{C_4} = \pm 1$ by the definition of $\chi$-group, i.e. $\chi_{C_4}(l)=\chi_{C_4}^*(r)=\chi_{C_4}^*(l) = \pm 1$, and therefore $\theta_{C_4} = \pm 1,\,\pm i$ from the second equation in Eq.\,\eqref{eq:wc4_relation}. Now, let us use the remaining $V$-ambiguity 

\begin{align}
	w_{C_4}(i) \rightarrow V(i) w_{C_4}(i) V^{-1}(C_4^{-1}(i)),
\end{align}
and set $w_{C_4}(r) = w_{C_4}(u) = w_{C_4}(d) = \mathbb{I}$ 
and $w_{C_4}(l) = \chi_{C_4} \mathbb{I}$ to satisfy Eq.\,\eqref{eq:wc4_relation}. Remaining $V$-ambiguity should satisfy

\begin{align}
	V(i) V^{-1}(C_4^{-1}(i)) = \mathbb{I}, \nonumber
\end{align}
and it indicates that only overall $V$-ambiguity is remained: $V(x,y,i) = V$. In the gauge we chosen so far, the gauge transformations associated with $T_1,\,T_2,\,C_4$ are 

\begin{align}
	& W_{T_{1,2}}(x,y,i) = \mathbb{I},\;\;\; \Theta_{T_{1,2}}(x,y)=1,\nn
	& W_{C_4}(x,y,i) = w_{C_4}(i),\;\;\; \Theta_{C_4}(x,y) = \theta_{C_4},
	\label{eq:wt12_wc4}
\end{align}
where

\begin{align}
	w_{C_4}(l) = \chi_{C_4} \mathbb{I},\;\;\;
	w_{C_4}(r/u/d) = \mathbb{I},\;\; \theta_{C_4} = \pm 1,\,\,\pm i,\nonumber
\end{align}
and $\chi_{C_4} = \pm 1$.

Let us consider the reflection symmetry\,($\sigma$). From the group relation $\sigma^{-1} T_2^{-1} \sigma T_1 = e$ and Eq.\,\eqref{eq:wt12_wc4}, we  obtain the following equation

\begin{align}
	& W_{\sigma}^{-1}(x,y,i) W_{\sigma}(T_2 x,y,i) = \chi_{T\sigma}^{(1)}(\sigma^{-1}(x,y,i)),\nn
	& \Theta_{\sigma}^*(x,y) \Theta_{\sigma}(T_2 x,y) = \prod_i [\chi_{T\sigma}^{(1)}(\sigma^{-1}(x,y,i))]^*.
	\label{eq:wtsigma1}
\end{align}
Using the $\varepsilon_{\sigma}$-ambiguity to set $\chi_{T\sigma}^{(1)}(x,y,i) = 1$, the remaining ambiguity should be $y$-independent: $\varepsilon_{\sigma}(x,y,i) = \varepsilon_{\sigma}(x,0,i) $\,(similar with Eq.\,\eqref{eq:ec4_x}). Then, RHS's of Eq.\,\eqref{eq:wtsigma1} are 1 so that we can obtain

\begin{align}
	W_{\sigma}(x,y,i) = W_{\sigma}(x,0,i),\;\;\;\;\;
	\Theta_{\sigma}(x,y) = \Theta_{\sigma}(x,0).	
\end{align}
Taking into account Eq.\,\eqref{eq:wt12_wc4}, the group relation $\sigma^{-1} T_2^{-1} \sigma T_1 = e$ gives us

\begin{align}
	& W_{\sigma}^{-1}(x,0,i) W_{\sigma}(T_1 (x,0,i)) = \chi_{T\sigma}^{(2)} (\sigma^{-1}(x,0,i)),\nn
	& \Theta_{\sigma}^*(x,0) \Theta_{\sigma}(T_1 (x,0)) = \prod_i [\chi_{T\sigma}^{(2)} (\sigma^{-1}(x,0,i))]^*.
	\label{eq:wtsigma2}
\end{align}
Since the remaining $\varepsilon_{\sigma}$ is also $y$-independent, we can use it to set $\chi_{T\sigma}^{(2)} (\sigma^{-1}(x,0,i)) = 1$, and site-independent $\varepsilon_{\sigma}$-ambiguity is remained: $\varepsilon_{\sigma}(x,y,i) = \varepsilon_{\sigma}(0,0,i)$. Consequently, Eq.\,\eqref{eq:wtsigma2} reads

\begin{align}
	& W_{\sigma}(x,y,i) = W_{\sigma}(0,0,i) \equiv w_{\sigma}(i),\nn
	& \Theta_{\sigma}(x,y) = \Theta_{\sigma}(0,0) \equiv \theta_{\sigma}.
\end{align}
From the group relation $\sigma^2=e$, constraints on $w_{\sigma}(i)$ and $\theta_{\sigma}$ are given such that $\theta_{\sigma} = \pm 1$ and 
\begin{align}
	w_{\sigma}(i) w_{\sigma}(\sigma(i)) = \chi_{\sigma}(i).
	\label{eq:const_wsigma}
\end{align}
Multiplying $w_{\sigma}^{-1}(i)$ from left and $w_{\sigma}(i)$ from right of Eq.\,\eqref{eq:const_wsigma} in sequence, we would obtain

\begin{align}
	w_{\sigma}(\sigma(i)) w_{\sigma}(i) = \chi_{\sigma}(i),
	\nonumber
\end{align}
while acting $\sigma$ on Eq.\,\eqref{eq:const_wsigma} gives

\begin{align}
	w_{\sigma}(\sigma(i)) w_{\sigma}(i) = \chi_{\sigma}(\sigma(i)),
	\nonumber
\end{align}
revealing $\chi_{\sigma}(\sigma(i)) = \chi_{\sigma}(i)$. This allows us to use the remaining $\varepsilon_{\sigma}$-ambiguity \,[$w_{\sigma}(i) \rightarrow \varepsilon_{\sigma}(i) w_{\sigma}(i)$ in Eq.\,\eqref{eq:const_wsigma} ]

\begin{align}
	\chi_{\sigma}(i) \rightarrow \varepsilon_{\sigma}(i) \varepsilon_{\sigma}(\sigma(i)) \chi_{\sigma}(i),\nonumber
\end{align}
and we set $\chi_{\sigma}(i) = 1$ for all legs\,($i$). Now, remaining $\varepsilon_{\sigma}$-ambiguity should satisfy

\begin{align}
	\varepsilon_{\sigma}(d) = \varepsilon^*(l) = \varepsilon_{\sigma}(r) = \varepsilon_{\sigma}^*(u).
	\label{eq:esigma}
\end{align}
The group relation $\sigma^{-1} C_4 \sigma C_4  = e$ and Eq.\,\eqref{eq:wt12_wc4} would give the following equations

\begin{align}
	w_{\sigma}^{-1}(\sigma(i)) w_{C_4}(\sigma(i)) w_{\sigma}(C_4^{-1} \sigma(i)) w_{C_4}(C_4(i)) = \chi_{C_4 \sigma}(i),
	\label{eq:wc4sigma}
\end{align}
and $\theta_{C_6}^2 = 1$ indicating that the group relation $\sigma^{-1} C_4 \sigma C_4  = e$ reduces the degree of freedom of $\theta_{C_4}$ from 4 to 2, i.e. $\theta_{C_4} = \pm 1,\,\pm i \rightarrow \theta_{C_4}=\pm 1$. Using the remaining $\varepsilon_{\sigma}$-ambiguity in Eq.\,\eqref{eq:wc4sigma}, we would obtain

\begin{align}
	\chi_{C_4\sigma}(i) \rightarrow \varepsilon_{\sigma}^*(\sigma(i)) \varepsilon_{\sigma}(C_4\sigma(i)) \chi_{C_4\sigma}(i).\nonumber
\end{align}
When $i=l$, the above equation gives

\begin{align}
	 \chi_{C_4\sigma}(l) \rightarrow & \;\varepsilon_{\sigma}^*(d) \varepsilon_{\sigma}(r) \chi_{C_4\sigma}(l)\nn
	&= \varepsilon_{\sigma}(l) \varepsilon_{\sigma}(r) \chi_{C_4\sigma}(l)
	=  \chi_{C_4\sigma}(l),\nonumber
\end{align} 
where Eq.\,\eqref{eq:esigma} and the definition of $\chi$-group are used for the first and second equality, respectively. Above equation implies that $\chi_{\sigma}(l)$ cannot be tuned by the $\varepsilon_{\sigma}$-ambiguity. However, as for $i=d$, $\chi_{\sigma}(d)$ is transformed as follows

\begin{align}
	\chi_{C_4\sigma}(d) \rightarrow \;\varepsilon_{\sigma}^*(l) \varepsilon_{\sigma}(d) \chi_{C_4\sigma}(d)
	= (\varepsilon_{\sigma}(l))^2 \chi_{C_4\sigma}(l),
	\nonumber
\end{align}
where Eq.\,\eqref{eq:esigma} is used for the equality, and we can tune $\varepsilon_{\sigma}(l)$ to set $\chi_{C_4\sigma}(d)=1$. Similarly, we can set $\chi_{C_4\sigma}(u)=1$, but $\chi_{C_4\sigma}(r)$ can not be fixed by $\varepsilon_{\sigma}$-ambiguity. Now, we define $\chi_{C_4\sigma}(l) = \chi_{C_4\sigma}^*(r) \equiv \chi_{C_4\sigma}$, and the remaining $\varepsilon_{\sigma}$-ambiguity should satisfy $\varepsilon_{\sigma} = \pm 1$ to maintain $\chi_{C_4\sigma}(u) = \chi_{C_4\sigma}^*(d) = 1$. Let us see the constraint Eq.\,\eqref{eq:const_wsigma} for each legs

\begin{align}
	& w_{\sigma}^{-1}(d) w_{C_4}(d) w_{\sigma}(l) w_{C_4}(d) = \chi_{C_4 \sigma}(l) \mathbb{I} = \chi_{C_4 \sigma} \mathbb{I},\nn
	& w_{\sigma}^{-1}(l) w_{C_4}(l) w_{\sigma}(u) w_{C_4}(r) = \chi_{C_4 \sigma}(d) = \mathbb{I},\nn
	& w_{\sigma}^{-1}(u) w_{C_4}(u) w_{\sigma}(r) w_{C_4}(u) = \chi_{C_4 \sigma}(r) \mathbb{I} = \chi_{C_4 \sigma}^* \mathbb{I},\nn
	& w_{\sigma}^{-1}(r) w_{C_4}(r) w_{\sigma}(d) w_{C_4}(l) = \chi_{C_4 \sigma}(u) = \mathbb{I}.
	\label{eq:wc4sigma1}
\end{align}
Acting $C_4$ on the firs equation of Eq.\,\eqref{eq:wc4sigma1} and comparing to the last equation of Eq.\,\eqref{eq:wc4sigma1}, we find 

\begin{align}
	w_{\sigma}^{-1}(r) w_{\sigma}(d) = \mathbb{I} = \chi_{C_4}^* \mathbb{I}, \nonumber
\end{align}
and therefore $\chi_{C_4} = 1$, the degree of freedom of $\chi_{C_4}$ is reduced from 2 to 1: $ \chi_{C_4} = \pm 1 \rightarrow \;\chi_{C_4} = 1$. Similarly, acting $C_4$ on the second equation of Eq.\,\eqref{eq:wc4sigma1} and comparing the first equation of Eq.\,\eqref{eq:wc4sigma1}, we find 

\begin{align}
	w_{\sigma}^{-1}(d) w_{\sigma}(l) = \chi_{C_4 \sigma} \mathbb{I} = \chi_{C_4 \sigma}^* \mathbb{I}, \nonumber
\end{align}
and therefore $\chi_{C_4\sigma}=1$. Inserting $\chi_{C_4} = \chi_{C_4 \sigma}=1$ back into Eqs. \eqref{eq:wt12_wc4} and \,\eqref{eq:wc4sigma1}, we can summarize the gauge transformation associated with the spatial symmetry operation as follows:

\begin{align}
	& W_{T_{1,2}}(x,y,i) = \mathbb{I},\;\;\; \Theta_{T_{1,2}}(x,y) = 1,\nn
	& W_{C_4}(x,y,i) = \mathbb{I},\;\;\; \Theta_{C_4}(x,y) = \theta_{C_4},\nn
	& W_{\sigma}(x,y,i) = w_{\sigma},\;\;\; \Theta_{\sigma}(x,y) = \theta_{\sigma},
	\label{eq:sol_space_group}
\end{align}
where 

\begin{align}
	\theta_{C_4} = \pm 1,\;\;\; \theta_{\sigma} = \pm 1,\;\;\; (w_{\sigma})^2 = \mathbb{I}. \nonumber
\end{align}

Let us take into account the time reversal symmetry\,($\mathcal{T}$). For the group relations $T_1^{-1} \mathcal{T}^{-1} T_1 \mathcal{T} = e$ and $T_2^{-1} \mathcal{T}^{-1} T_2 \mathcal{T} = e$, we can follow exactly the same procedure for $T_1^{-1} \sigma^{-1} T_1 \sigma = e$ and $T_2^{-1} \sigma^{-1} T_2 \sigma = e$ such that in the gauge we chosen 

\begin{align}
	W_{\mathcal{T}}(x,y,i) = w_{\mathcal{T}}(i),\;\;\; \Theta_{\mathcal{T}}(x,y) = \theta_{\mathcal{T}},
	\label{eq:wtime12}
\end{align}
and the remaining $\varepsilon_{\mathcal{T}}$-ambiguity should satisfy $\varepsilon_{\mathcal{T}}(x,y,i) = \varepsilon_{\mathcal{T}}(i)$. Taking into account Eqs.\,\eqref{eq:sol_space_group}, \eqref{eq:wtime12}, from the group relation $C_4^{-1} \mathcal{T}^{-1} C_4 \mathcal{T} = e $, we can find

\begin{align}
	[w_{\mathcal{T}}^{-1}(C_4(i))]^* w_{\mathcal{T}}^*(i) = \chi_{C_4\mathcal{T}}(i), \nonumber
\end{align}
and the remaining $\varepsilon_{\mathcal{T}}$-ambiguity can be used to set $\chi_{C_4\mathcal{T}}(i)=1$ implying that $w_{\mathcal{T}}(i)$ does not depend on the leg index anymore:

\begin{align}
	w_{\mathcal{T}}(i) = w_{\mathcal{T}}(C_4(i)) = w_{\mathcal{T}}(l) \equiv w_{\mathcal{T}}. \nonumber
\end{align}
Now, we used the remaining $\Phi$-ambiguity to set

\begin{align}
	\theta_{\mathcal{T}} \rightarrow \Phi \theta_{\mathcal{T}} (\mathcal{T} \circ \Phi^*) = \Phi^2 \theta_{\mathcal{T}} = 1,\nonumber
\end{align}
and therefore

\begin{align}
	W_{\mathcal{T}}(x,y,i) = w_{\mathcal{T}},\;\;\; \Theta_{\mathcal{T}}(x,y) = 1.
	\label{eq:wtimec4}
\end{align}
In the gauge we chosen so far, the group relations $\sigma^{-1} \mathcal{T}^{-1} \sigma \mathcal{T} = e$ and $\mathcal{T}^2 = e$ gives us following constraints

\begin{align}
	& w_{\sigma}^{-1} [w_{\mathcal{T}}^{-1}]^* w_{\sigma}^* w_{\mathcal{T}}^* = \chi_{\sigma\mathcal{T}}, \nn
	&  w_{\mathcal{T}} w_{\mathcal{T}}^* = \chi_{\mathcal{T}},
\end{align}
where $\chi_{\sigma\mathcal{T}} = \pm 1$ and $\chi_{\mathcal{T}} = \pm 1$. 

Finally, let us incorporate the SU(2) spin rotation symmetry. Since $W_{T_{1,2}}$, $W_{C_4}$ are just identity\,[Eq.\,\eqref{eq:sol_space_group}], the group relations $g^{-1} U_{\theta\vec{n}}^{-1} g U_{\theta\vec{n}} = e$, where $g = T_1,\,T_2,\,C_4$, simply give

\begin{align}
	W_{\theta}^{-1}(g(x,y,i)) W_{\theta}(x,y,i) = \chi_{g\theta}(x,y,i).\nonumber
\end{align}
Here, $W_{\theta}$ is a gauge transformation associated with time-reversal symmetry operation forming a SU(2) representation\,\cite{vidal12}, and therefore LHS of above equation forms an 1D representation of SU(2) symmetry as RHS is able to form only an 1D representation\,(U(1)). Due to the fact that there is no non-trivial 1D representation of SU(2) symmetry, we can set $\chi_{g\theta}(x,y,i)=1$ to be unity. Similarly, the phase factor $\Theta_{\theta}(x,y)$ associated with $W_{\theta}$ should form an 1D representation of SU(2) symmetry itself, and therefore $\Theta_{\theta}(x,y) = 1$. Then, it is straightforward to prove from above equation that $W_{\theta}$ is independent on coordinate and leg index in the gauge we chosen, i.e.

\begin{align}
	W_{\theta}(x,y,i) \equiv w_{\theta}. \label{eq:w_theta}
\end{align}
Here, we consider a group relation $U_{2\pi \vec{n}} = e$ giving us a relation

\begin{align}
	W_{\theta=2\pi}(x,y,i) = w_{\theta=2\pi} = \chi_{\theta=2\pi}(x,y,i) = \chi_{\theta=2\pi} = \pm 1, \label{eq:2pi_rotate}
\end{align}
where we use the fact that $w_{\theta}$ is site and leg independent \,[Eq.\,\eqref{eq:w_theta}] for the last equality. 

From the group relations $\sigma^{-1} U_{\theta \vec{n}}^{-1} \sigma U_{\theta \vec{n}} = e$ and $\mathcal{T}^{-1} U_{\theta \vec{n}}^{-1} \mathcal{T} U_{\theta \vec{n}} = e$, we obtain the following constraint

\begin{align}
	& w_{\sigma}^{-1} w_{\theta}^{-1} w_{\sigma} w_{\theta} = \chi_{\sigma\theta},\nn
	& w_{\mathcal{T}}^{-1} [w_{\theta}^{-1}]^* [w_{\mathcal{T}}]^* [w_{\theta}]^* = \chi_{\mathcal{T}\theta},\nonumber
\end{align}
where $\chi_{\sigma\theta} = \pm 1$ and $\chi_{\mathcal{T}\theta} = \pm 1$. Notice that $w_{\theta}$ forms a {\it continuous} group while both $\chi_{\sigma \theta}$ and $\chi_{\mathcal{T}\theta}$ have only two {\it discrete} values, which seems impossible from above equation and suggests that $\chi_{\sigma \theta}$ and $\chi_{\mathcal{T} \theta}$ should be fixed to a proper value $+1$ or $-1$. When $\theta=0$ or $U_{\theta=0} = \mathbb{I}$\,(no spin rotation), $w_{\theta=0}$ must be identity, and therefore it is easy to see $\chi_{\sigma \theta=0} = +1 = \chi_{\mathcal{T} \theta=0}$ from above equation. Consequently, we have to fix $\chi_{\sigma \theta} = +1 = \chi_{\mathcal{T}\theta}$ for all $\theta$, and constraint above is recast as

\begin{align}
	& w_{\sigma}^{-1} w_{\theta}^{-1} w_{\sigma} w_{\theta} = \mathbb{I},\nn
	& w_{\mathcal{T}}^{-1} [w_{\theta}^{-1}]^* [w_{\mathcal{T}}]^* [w_{\theta}]^* = \mathbb{I}.
	\label{eq:wsigmatime_const}
\end{align}
In summary,

\begin{align}
	& W_{T_{1,2}}(x,y,i) = \mathbb{I},\;\;\; \Theta_{T_{1,2}}(x,y) = 1,\nn
	& W_{C_4}(x,y,i) = \mathbb{I},\;\;\; \Theta_{C_4}(x,y) = \theta_{C_4},\nn
	& W_{\sigma}(x,y,i) = w_{\sigma},\;\;\; \Theta_{\sigma}(x,y) = \theta_{\sigma},\nn
	& W_{\mathcal{T}}(x,y,i) = w_{\mathcal{T}},\;\;\; \Theta_{\mathcal{T}}(x,y) = 1,\nn
	& W_{\theta}(x,y,i) = w_{\theta},\;\;\; \Theta_{\theta}(x,y) = 1,
	\label{eq:sol_space_group}
\end{align}
where 

\begin{align}
	\theta_{C_4} = \pm 1,\;\;\; \theta_{\sigma} = \pm 1,. \nonumber
\end{align}
and constraints for unknown $w_g$'s are

\begin{align}
	& (w_{\sigma})^2 = \mathbb{I}, \nn
	& w_{\sigma}^{-1} w_{\theta}^{-1} w_{\sigma} w_{\theta} = \mathbb{I},\nn
	& w_{\mathcal{T}}^{-1} [w_{\theta}^{-1}]^* [w_{\mathcal{T}}]^* [w_{\theta}]^* = \mathbb{I}.
	\nonumber
\end{align}
%


Let us find specific form of the unknown $w_g$'s. First, we can use the remaining $V$-ambiguity\,[$V(x,y,i)=V$] to have

\begin{align}
	w_{\theta \vec{n}} = \bigoplus_{i=1}^M \left( I_{d_i} \otimes e^{i\theta \vec{n} \cdot \vec{S}_i} \right),
	\label{eq:w_theta_sol}
\end{align}
where $M$ is the number of different spins\,($\vec{S}_i$) living on the virtual legs and $d_i$ is the dimension of degeneracy flavor space implying that the bond dimension is $D=\sum_{i=1}^M d_i (2 S_i + 1)$. Notice that $w_{\theta \vec{n}}$ is nothing but the SU(2) spin rotation operator on virtual spins. Here, Eq.\,\eqref{eq:2pi_rotate} gives us a constraint that we can assign only integer\,(half-integer) spins on virtual legs for $\chi_{\theta=2\pi} = +1\,(-1)$. Therefore, the mixture of half-integer and integer spins on virtual legs immediately breaks SU(2) spin rotation symmetry in the gauge we chosen. Remaining $V$-ambiguity should leave $w_{\theta \vec{n}}$ invariant, i.e. $V^{-1} w_{\theta \vec{n}} V = w_{\theta \vec{n}}$, and therefore 

\begin{align}
	V = \bigoplus_{i=1}^M \left( \widetilde{V}_i \otimes \mathbb{I}_{2S_i+1} \right),
\end{align}
where $V_i$ is a $d_i$-dimensional invertible matrix. 

Next, the constraint $w_{\theta}^{-1} (w_{\mathcal{T}}^{-1})^* w_{\theta}^* w_{\mathcal{T}}^* = \mathbb{I}$ can be recast as $w_{\theta} w_{\mathcal{T}} = w_{\mathcal{T}} w_{\theta}^*$, and it indicates that $w_{\mathcal{T}}$ reverses spin as follows $S_x \rightarrow -S_x$, $S_y \rightarrow S_y$, $S_z \rightarrow -S_z$[Eq.\,\eqref{eq:w_theta_sol}]. Therefore, using the remaining $V$-ambiguity, we can set $w_{\mathcal{T}}$ as follows
\begin{align}
	w_{\mathcal{T}} = \bigoplus_{i=1}^M \left( \widetilde{w}_{\mathcal{T}}^i \otimes e^{i\pi S_y} \right),
	\label{eq:w_time}
\end{align}
where $\widetilde{w}_{\mathcal{T}}^i$ is a $d_i$-dimensional invertible matrix. Now, the constraint $w_{\mathcal{T}} w_{\mathcal{T}}^* = \mathbb{I}$ is recast as 

\begin{align}
	\bigoplus_{i=1}^M [\widetilde{w}_{\mathcal{T}}^i (\widetilde{w}_{\mathcal{T}}^i)^* ] \otimes [e^{i\pi S_i^y} e^{-i(\pi S_i^y)^*}] = \chi_{\mathcal{T}}\mathbb{I}, \nonumber
\end{align}
where $e^{i\pi S_i^y} e^{i\pi S_i^y}= \chi_{\theta=2\pi} \mathbb{I}_{2S_i+1}$, and therefore $\widetilde{w}_{\mathcal{T}}^i (\widetilde{w}_{\mathcal{T}}^i)^* = \chi_{\mathcal{T}} \, \chi_{\theta=2\pi} \mathbb{I}_{d_i}$ to satisfy above equation. Physical meaning of $w_{\mathcal{T}} w_{\mathcal{T}}^*$ is to act time-reversal operation twice on Hilbert space of virtual leg. Similarly, $\widetilde{w}_{\mathcal{T}}^i (\widetilde{w}_{\mathcal{T}}^i)^*$ reverses the time of degeneracy flavor space twice, and therefore $\widetilde{w}_{\mathcal{T}}^i (\widetilde{w}_{\mathcal{T}}^i)^* = + \mathbb{I}_{d_i}$\,($\chi_{\theta=2\pi} = \chi_{\mathcal{T}}$) means that the degeneracy flavor space of virtual legs should be Kramers singlet while Kramers double for $\widetilde{w}_{\mathcal{T}}^i (\widetilde{w}_{\mathcal{T}}^i)^* = - \mathbb{I}_{d_i}$\,($\chi_{\theta=2\pi} \neq \chi_{\mathcal{T}}$). Consequently, only 4 different types of Hilbert space of virtual legs are allowed and it is summarized in Table\,\ref{table:virtual_legs}. When the degeneracy flavor space is Kramers singlet\,($\chi_{\mathcal{T}}=+1$), by choosing proper basis  we can set $\widetilde{w}_{\mathcal{T}}^i = \mathbb{I}_{d_i}$ and the remaining $V$-ambiguity should satisfy $\widetilde{V}_i\,\mathbb{I}_{d_i}\,(\widetilde{V}_i^{-1})^* = \mathbb{I}_{d_i}$ such that $\widetilde{V}_i$ is a real matrix\,($\widetilde{V}_i = \widetilde{V}_i^*$). However, when Kramers doublet is considered as degeneracy flavor space\,($\chi_{\mathcal{T}}=-1$), we should set $\widetilde{w}_{\mathcal{T}}^i$ to be antisymmetric: $\widetilde{w}_{\mathcal{T}}^i = \Omega_i = i\sigma_y \otimes \mathbb{I}_{d_i/2}$. In this case, the remaining $V$-ambiguity should satisfy $\widetilde{V}_i \, \Omega_i (\widetilde{V}_i^{-1})^* = \Omega_i$.

\begin{center}
  \begin{table}
    \begin{tabular}{| c || c |}
      \hline
      ( $\chi_{\mathcal{T}}$, $\chi_{\theta=2\pi}$) & degeneracy flavor space $\otimes$ spin space \\
      \hline \hline
      (+1,+1) & Kramers singlet $\otimes$ integer spins \\ \hline
      (+1,-1) & Kramers singlet $\otimes$ half-integer spins \\ \hline
      (-1,+1) & Kramers doublet $\otimes$ integer spins \\ \hline
      (-1,+1) & Kramers doublet $\otimes$ half-integer spins \\ \hline
      \end{tabular}
    \caption[Table caption text]{4 different types of Hilbert space of virtual legs in terms of ($\chi_{\mathcal{T}}, \chi_{\theta=2\pi}$).}
    \label{table:virtual_legs}
  \end{table}
\end{center}

In the same manner, using the constraint $w_{\sigma}^{-1} w_{\theta}^{-1} w_{\sigma} w_{\theta} = \mathbb{I}$, we can set

\begin{align}
	w_{\sigma} = \bigoplus_{i=1}^M \left( \widetilde{w}_{\sigma}^i \otimes \mathbb{I}_{2S_i+1} \right),
	\label{eq:w_sigma}
\end{align}
where $\widetilde{w}_{\sigma}^i$ is a $d_i$-dimensional matrix satisfying a condition $ (\widetilde{w}_{\sigma}^i)^* = \chi_{\sigma\mathcal{T}} \widetilde{w}_{\sigma}^i$ followed from the constraint $w_{\sigma}^{-1} (w_{\mathcal{T}}^{-1})^* w_{\sigma}^* w_{\mathcal{T}}^* = \chi_{\sigma \mathcal{T}}\mathbb{I}$. Therefore, $\widetilde{w}_{\sigma}^i$ is a real matrix for $\chi_{\sigma\mathcal{T}} = +1$ while an imaginary matrix for $\chi_{\sigma\mathcal{T}} = -1$. Further, we can find $(\widetilde{w}_{\sigma}^i)^2 = \mathbb{I}_{d_i}$ from the constraint $(w_{\sigma})^2 = \mathbb{I}$.

\subsection{Constraints on bond tensor}

In this subsection, we find some constraints on bond tensor using the gauge transformations\,[Eqs.\,\eqref{eq:sol_space_group}, \eqref{eq:w_theta_sol}, \eqref{eq:w_time} and \eqref{eq:w_sigma} ] and some specific form of bond tensor allowed in the gauge we chosen. 

In symmetric PEPS, the bond tensor follows Eq.\,\eqref{eq:sym_peps} for a given symmetry operation $R$. Acting inverse of $R$, we have

\begin{align}
	R^{-1} \circ \mathbf{B} = R^{-1} W_{R} R \circ \mathbf{B}, \nonumber
\end{align} 
and its matrix representation is the following\,\cite{shenghan15A}

\begin{widetext}
\begin{align}
	[\mathbf{B}(R(x,y,i);R(x',y',i'))]_{\alpha\beta} & = [\{W_R^{-1}(R(x,t,i))\}^t]_{\alpha\alpha'} [\{W_R^{-1}(R(x',t',i'))\}^t]_{\beta\beta'} [\mathbf{B}(x,y,i;x',y',i')]_{\alpha'\beta'}\nn
	& = [W_R^*(R(x,t,i))]_{\alpha\alpha'} [\mathbf{B}(x,y,i;x',y',i')]_{\alpha'\beta'} [ W_R^{-1}(R(x',t',i'))]_{\beta'\beta} ,
\end{align}
\end{widetext}
where $(x,y,i;x',y',i')$ is the position of bond tensor connecting two legs $i$ and $i'$, $t$ denotes the transpose of matrix, and we used the fact that all $W_R$'s are unitary in the second equality. First, using $W_{T_{1,2}} = W_{C_4} = \mathbb{I}$, we can find that the bond tensor is independent on the site and leg indices: $\mathbf{B}(x,y,i;x',y',i') = \mathbf{B}$. From the spin rotation symmetry $R=U_{\theta\vec{n}}$, we have the constraint $\mathbf{B} = W_{\theta}^* \mathbf{B} W_{\theta}^{-1}$. Its physical meaning is that action of arbitrary spin rotation on Hilbert space of virtual spins remains itself, which implies that bond tensor must be spin singlet in the gauge we chosen. Therefore, we can set the bond tensor as follows

\begin{align}
	\mathbf{B} = \bigoplus_{i=1}^M \left( \widetilde{B}_i \otimes K_i \right),
\end{align}
where $\widetilde{B}_i$ is a $d_i$-dimensional matrix characterizing the degeneracy flavor space and $K_i$ is a $(2S_i+1)$- dimensional matrix standing for singlet state. In other words, $\hat{K}_i = \langle S_i, m_{\alpha} ; S_i, m_\beta | (K_i)_{\alpha\beta}$ is a singlet state: 

\begin{align}
	\langle S_i, m_{\alpha} ; S_i, m_\beta | (K_i)_{\alpha\beta} (S_i^{\rm tot})^2 = 0,
	\nonumber
\end{align}
where $S_i^{\rm tot} = S_i \otimes \mathbb{I} + \mathbb{I} \otimes S_i$. The most general form of $K_i$ satisfying above equation can be obtained from Clebsch-Gordan\,(CG) coefficients such that

\begin{align}
	(K_i)_{\alpha\beta} = \sqrt{2S_i +1}C_{S_i,
	m_\alpha; S_i, m_{\beta}}^{0,0} 
	= (-1)^{S_i-m_{\alpha}} \delta_{m_{\alpha},-m_{\beta}},
\end{align}
where $C_{S_i, m_\alpha; S_i, m_{\beta}}^{0,0}$ is CG tensor for fusing two spins into singlet, and $m_{\alpha} = -S_i+\alpha-1$ is the $S_z$ quantum number. From the time reversal symmetry $R = \mathcal{T}$, we obtain 

\begin{align}
	\mathbf{B}^* = W_{\mathcal{T}}\,\mathbf{B}\,W_{\mathcal{T}}^t
	& = \bigoplus_{i=1}^M \left\{ [ \widetilde{w}_{\mathcal{T}}^i\,\widetilde{B}_i (\widetilde{w}_{\mathcal{T}}^i)^t] \otimes K_i \right\}, \nonumber
\end{align}
and therefore

\begin{align}
	\widetilde{B}_i^* = \widetilde{w}_{\mathcal{T}}^i\,\widetilde{B}_i (\widetilde{w}_{\mathcal{T}}^i)^t.
\end{align}
It can be easily shown that $\widetilde{B}_i$ should be real matrix for $\chi_{\mathcal{T}} = +1$ while $\widetilde{B}_i^* = \Omega_i\, \widetilde{B}_i\, (\Omega_i)^t$ for $\chi_{\mathcal{T}} = -1$ using the fact that $\widetilde{w}_{\mathcal{T}}^i = \mathbb{I}_{d_i}\,(\Omega_i)$ for $\chi_{\mathcal{T}} = +1\,(-1)$. In similar way, we can find the last constraint on bond tensor from the reflection symmetry: 
\begin{align}
	\widetilde{B} = \chi_{\sigma\mathcal{T}} \widetilde{w}_{\sigma}^i\, \widetilde{B}_i\, \widetilde{w}_{\sigma}^i.
\end{align}
\subsection{Constraints on site tensor}

In the process of fixing gauge, we have already shown that the site tensor is site-independent from the translational symmetry\,($T_{1,2}$). 

Let us define the Hilbert space of site tensor\,($\mathbb{V}_{\mathbf{T}}$). It can be obtained by tensor product of Hilbert spaces of physical spin and virtual spins  as follows

\begin{align}
	\mathbb{V}_{\mathbf{T}} & \cong \mathbb{V}_p \otimes \mathbb{V}_l \otimes \mathbb{V}_r \otimes \mathbb{V}_u \otimes \mathbb{V}_d\nn
	& \cong \bigoplus_{i_l,\cdots,i_d = 1}^M \left( \mathbb{D}_{i_l i_r i_u i_d}\otimes \mathbb{V}_{p, i_l, i_r, i_u, i_d}^{S=0} \otimes \mathbb{V}_{S=0} \right),
\end{align}
where $p$ denotes the physical spin, $\mathbb{D}_{i_l i_r i_u i_d }$ denotes the total degeneracy flavor space and $\mathbb{V}_{i_l i_r i_u i_d }^{S=0}$ is the fusion space describing how to fuse the physical and virtual spins to construct singlet state. Considering SU(2) spin rotation symmetry, general expression of the site tensor can be set to be

\begin{align}
	\mathbf{T} = \bigoplus_{i_l,\cdots,i_d, \mu} \left( \widetilde{T}_{i_l, i_r, i_u, i_d}^{\mu} \otimes K_{p,i_l,i_r,i_u,i_d}^{\mu}\right),
\end{align}
where $\mu$ labels different way to fuse spins into singlet state.

\bibliographystyle{apsrev}
\bibliography{reference.bib}

\end{document}